\documentclass[11pt,twoside,preprint]{article}
\usepackage[abbrvbib]{jmlr2e}
\jmlrheading{25}{2024}{1-X}{09/24}{X}{00-000}{Matteo Bettini, Ajay Shankar and Amanda Prorok} 
\ShortHeadings{System Neural Diversity}{Bettini, Shankar, Prorok}
\firstpageno{1} 

\usepackage{multicol}
\usepackage{booktabs} 
\usepackage{siunitx, amsmath, amsfonts, amssymb}
\usepackage[super]{nth}
\usepackage[export]{adjustbox}
\usepackage[font=footnotesize]{caption}
\usepackage[font=footnotesize]{subcaption}
\usepackage{xcolor}
\usepackage[normalem]{ulem}

\graphicspath{{figures/}}
\newtheorem{property}{Property}
\newtheorem{define}{Definition}

\newcommand{\R}{\mathbb{R}}
\renewcommand{\cite}{\citep}

\begin{document}

\title{System Neural Diversity: Measuring Behavioral Heterogeneity in Multi-Agent Learning}

\author{%
\name Matteo Bettini \email mb2389@cl.cam.ac.uk \\
\addr Department of Computer Science and Technology,\\ University of Cambridge, UK
    \AND
\name {Ajay Shankar} \email {as3233@cl.cam.ac.uk}\\
\addr Department of Computer Science and Technology,\\ University of Cambridge, UK
    \AND
\name {Amanda Prorok} \email {asp45@cl.cam.ac.uk}\\
\addr Department of Computer Science and Technology,\\ University of Cambridge, UK
}

\editor {NA}

\maketitle

\begin{abstract}%
Evolutionary science provides evidence that diversity confers resilience in natural systems. Yet, traditional multi-agent reinforcement learning techniques commonly enforce homogeneity to increase training sample efficiency. When a system of learning agents is not constrained to homogeneous policies, individuals may develop diverse behaviors, resulting in emergent complementarity that benefits the system. Despite this, there is a surprising lack of tools that \textit{quantify} behavioral diversity. Such techniques would pave the way towards understanding the impact of diversity in collective artificial intelligence and enabling its control. In this paper, we introduce System Neural Diversity (SND): a measure of behavioral heterogeneity in multi-agent systems. We discuss and prove its theoretical properties, and compare it with alternate, state-of-the-art behavioral diversity metrics used in the robotics domain. Through simulations of a variety of cooperative multi-robot tasks, we show how our metric constitutes an important tool that enables measurement and control of behavioral heterogeneity. In dynamic tasks, where the problem is affected by repeated disturbances during training, we show that SND allows us to measure latent resilience skills acquired by the agents, while other proxies, such as task performance (reward), fail to. Finally, we show how the metric can be employed to control diversity, allowing us to enforce a desired heterogeneity set-point or range. We demonstrate how this paradigm can be used to bootstrap the exploration phase, finding optimal policies faster, thus enabling novel and more efficient MARL paradigms.
\end{abstract}

\begin{keywords}
  Diversity, Multi-Agent Reinforcement Learning, Heterogeneity Measure
\end{keywords}

\section{Introduction}

Diversity is key to collective intelligence~\cite{woolley2015collective} and commonplace in natural systems
~\cite{kellert1997value}. 
Just as biologists and ecologists have demonstrated the role of functional diversity in ecosystem survival~\cite{cadotte2011beyond}, it has also been shown to provide resilience and performance benefits in Multi-Agent Reinforcement Learning (MARL)~\cite{bettini2023hetgppo}.
While a variety of measures of diversity exist for natural systems~\cite{tucker2017guide}, there is a lack of work studying \textit{behavioral} diversity\footnote{also referred to as `neural' diversity} in engineered systems.
Developing a principled diversity measure would allow us to directly quantify previously unobservable properties of the system (such as resilience) as well as enable its control (\textit{e.g.}, in a closed-loop fashion).
In this research, we are interested in measuring this diversity and its impact in multi-robot tasks, where collective intelligence and cooperation are critical to success. 

In order to boost sample efficiency, traditional MARL algorithms constrain policies to be identical for agents with the same objective~\cite{gupta2017cooperative,rashid2018qmix,sukhbaatar2016learning}. This causes the agents to become behaviorally homogeneous. On the other hand, when policies can diverge, and thus produce different action distributions for the same observation, we refer to the team as `heterogeneous'.
Recent MARL paradigms for training heterogeneous policies have been shown to provide intrinsic resilience benefits~\cite{bettini2023hetgppo}.
While several recent methods promote behavioral diversity in MARL~\cite{jaderberg2019human, wang2021rode, christianos2021scaling, chenghao2021celebrating, wang2020roma}, they lack a principled measure for it. 
Because of this, they are limited to analyzing the system from a performance (i.e., reward) perspective, which is frequently a suboptimal proxy for the underlying diversity.
The development of a principled behavioral heterogeneity metric would enable an analysis of the impact of diversity, and subsequently, its control~\cite{bettini2024controlling}.

In this paper, we introduce System Neural Diversity (SND), a measure of behavioral heterogeneity in multi-agent learning.
In addition to satisfying formal properties of a \textit{metric}, SND follows two key properties that are desirable in such a measure: (1) given a fixed inter-agent pairwise behavioral distance, the metric should not depend on the number of agents, and, (2) the metric should decrease as more agents assume the same behaviors.
Property (1) allows us to compare heterogeneity across different team sizes, while Property (2) allows the metric to measure behavioral redundancy.
To compute SND, we first define a pairwise inter-agent behavioral distance. 
Pairwise distances are then aggregated into a system-wide diversity metric.
SND is the first behavioral diversity metric that can be computed in closed-form for continuous stochastic action distributions. 
Furthermore, we juxtapose SND with Hierarchic Social Entropy (HSE)~\cite{balch2000hierarchic}, a state-of-the-art behavioral heterogeneity metric used in robotics, and show that HSE does not follow the properties defined, thus providing an alternate (and yet, complementary) view of behavioral diversity.

Using the proposed metric, we perform a set of evaluations to study the following research question: \textit{``What is the impact of neural diversity on multi-agent learning?''} Our studies in static (i.e., fixed during training) and dynamic (i.e., changing during training) cooperative multi-robot problems show that SND is a powerful tool to assess previously unobservable performance and resilience properties of multi-robot systems.
In particular, in cases where the task undergoes repeated dynamic disruptions, we show that measuring SND reveals \textit{latent} diversity in skills that the system has developed to cope with the disruptions.
We further show that
SND can be used to explicitly control for a target diversity during training, thus bootstrapping the search for the optimal policy and enabling the emergence of novel strategies.

A summary of our contributions in this work is as follows:
\begin{itemize}
    \item We introduce SND, a novel metric to quantify behavioral heterogeneity in policies: SND is currently the only diversity metric that can be computed in closed-form for continuous stochastic action distributions;
    \item We present and prove SND's theoretical properties, juxtaposing it to previous diversity measures;
    \item Through a variety of collective multi-robot tasks, we study the impact and insights gathered from measuring and controlling diversity in MARL, showing how SND provides a novel tool, enabling us to analyze as well as act upon previously unobservable latent properties in multi-agent learning.
\end{itemize}


\section{Related Work}
In this section, we give an overview of existing diversity measures used in the multi-agent and multi-robot domain. In \autoref{tab:related} we provide a detailed table comparing related works to ours, highlighting how SND is the only statistical metric that can be computed in closed-form between the stochastic continuous action distributions provided as output by the policies. \autoref{app:related} contains further discussion and comparisons on diversity indices in the biological domain.

\begin{table*}[t]
\caption{Comparison of SND with the related works.}
\label{tab:related}
\resizebox{\linewidth}{!}{%
\begin{tabular}{lccccc}
\toprule
Paper & System aggregation$^a$ & Behavioral distance$^b$ & Action space$^c$ & Probability distributions$^d$ & Population-based$^e$ \\
\midrule
SND (Ours) & Mean & Wasserstein& Continuous & Yes (closed-form) & No \\
\citet{mckee2022quantifying} & Mean & Total variation distance& Discrete & Yes (empirically estimated) & Yes \\
\citet{yu2021informative} & NA & KL & Discrete & Yes (empirically estimated) & Yes \\
\citet{yu2021informative} & NA & Wasserstein & Continuous & Yes (empirically estimated) & Yes \\
\citet{hu2022policy} & Mean & Symmetric KL & Discrete & Yes (closed-form) & No \\
\citet{balch2000hierarchic} & Clustering & Difference & Continuous & No (deterministic) & No \\
\citet{liu2021towards} & None & $f$-divergence & Discrete & Yes (closed-form) & Yes \\
\citet{parker2020effective} & Determinant & Kernel & Continuous & No (deterministic) & Yes (single-agent) \\
\bottomrule
\end{tabular}}\\

$^a$ Aggregation function to compute the system-level value\\
$^b$ Pairwise distance function used to compare agent pairs\\
$^c$ Action space type\\
$^d$ Whether the behavioral distance is computed between probability distributions\\
$^e$ If the approach is population-based\\
\end{table*}

\subsection{Diversity Measures in MARL}

In this work, we are interested in \textit{measuring} diversity, and thus, we only consider MARL policies that are already heterogeneous, i.e., those that do not employ parameter sharing.
This notion of diversity has gained increasing attention in recent years, and several works have outlined the caveats and implications of parameter sharing~\cite{christianos2021scaling,fu2022revisiting,bettini2023hetgppo}.
These have shown that even if homogeneous policies can emulate heterogeneous behavior in certain cases (using, for instance, some input context), they are often brittle in the presence of real-world noise, and sometimes, completely prevent task success.
We refer to \citet{bettini2023hetgppo} for a classification of existing heterogeneous MARL solutions. 
While some solutions are able to promote diversity to boost training performance~\cite{wang2021rode, christianos2021scaling, chenghao2021celebrating, wang2020roma,mahajan2019maven,jiang2021emergence,wang2019influence}, they do so without being able to measure the resulting heterogeneity -- the problem of developing a reliable diversity metric is frequently overlooked.

Nevertheless, some prior work has considered approximate diversity metrics.
In~\citet{mckee2022quantifying} the authors introduce a diversity metric that uses sampled discrete actions from agents' policies to approximate their distributions, and then total variation distance to compute the divergence among them. 
Similarly, in~\citet{yu2021informative}, behavioral distances are obtained using sampled action datapoints.
These methods use approximated distributions, leading to a decrease in the metric's accuracy. 
In contrast, SND measures the distance between agent action distributions in closed-form, avoiding approximations.
Another action-based diversity metric~\cite{hu2022policy}
uses symmetric Kullback–Leibler (KL) divergence to measure behavioral distances in discrete action spaces. However, symmetric KL does not satisfy the triangle inequality property, and hence, it is not a statistical metric. 
In~\citet{liu2021towards} the authors propose to use f-divergence between occupancy measures to quantify behavioral diversity in zero-sum games. The proposed solution, however, becomes computationally intractable for more general Markov games.

Population-based RL employs methods from Quality-Diversity optimization~\cite{chatzilygeroudis2021quality, stanley2019designing},
in which populations are `bred' to maximize both performance and diversity, with some techniques to analyze diversity in this domain~\cite{doncieux2010behavioral,masood2019diversity}. 
In~\citet{parker2020effective}, authors propose a diversity measure that computes the population diversity as the determinant of the agents' behavioral distance matrix. The agent distances, however, are bounded and are computed between deterministic actions over randomly sampled states, which can lead to policies being evaluated in states they have not been trained on.
In contrast, our work does not train populations of agents, and considers the problem of measuring the behavioral diversity of concurrently acting agents in a team.

\subsection{Diversity Measures in Robotics}
When agents are embodied as robots, physical differences can emerge, leading to different capabilities. \citet{prorok2017impact} characterize such differences in a diversity matrix representing the species and traits of a robot population. The rank of such matrices can be related to behavioral redundancy in the population and \textit{eigenspecies} can then be identified. 

Behavioral differences, however, are not always correlated with physical ones and need a dedicated measure in policy space. \citet{li2004learning} propose a behavioral specialization metric that is, however, obtained through a correlation with team performance. \citet{twu2014measure} propose a computation-light heterogeneity metric that assumes a fixed number of diversity classes.
Closest to our work, Balch's Hierarchic Social Entropy (HSE)~\cite{balch2000hierarchic}
computes the behavioral distance between two agents as the difference of their deterministic actions over all states.
It then uses this distance to compute a team level metric.
In particular, agents are hierarchically divided into behavioral clusters. For each hierarchical level, the Shannon entropy is computed on the distribution of agents in behavioral clusters. At hierarchical level $l$, for example, agents $i,j$ with behavioral distance $d_{\mathrm{HSE}}(i,j)\leq l$ will belong to the same cluster and $E(l)$ will denote the Shannon entropy of the clusters. HSE is then computed as
\begin{equation}
    \mathrm{HSE} = \int_0^\infty E(l) dl.
    \label{eq:hse}
\end{equation}
Since Shannon entropy does not take into account the distance between classes, hierarchical clustering is used in HSE to make the metric depend on such distances.
We will analyze and compare HSE with SND in the following sections, showing how these metrics provide complementary tools to assess diversity. 

\section{Problem Formulation}

\label{sec:formulation}

We now formulate the MARL problem analyzed in this work. To do so, we first introduce the multi-agent extension of a Partially Observable Markov Decision Process (POMDP)~\cite{kaelbling1998planning}.

\subsection{Partially Observable Markov Games}
A Partially Observable Markov Game (POMG) is defined as a tuple
$$\left \langle \mathcal{N}, \mathcal{S}, \left \{ \mathcal{O}_i \right \}_{i \in \mathcal{N}}, \left \{ \sigma_i \right \}_{i \in \mathcal{N}},  \left \{ \mathcal{A}_i \right \}_{i \in \mathcal{N}}, \left \{ \mathcal{R}_i \right \}_{i \in \mathcal{N}}, \mathcal{T}, \gamma \right \rangle,$$
where $\mathcal{N} = \{1,\ldots, n\}$ denotes the set of agents,
$\mathcal{S}$ is the state space, shared by all agents, and,
$\left \{ \mathcal{O}_i \right \}_{i \in \mathcal{N}}$ and
$\left \{ \mathcal{A}_i \right \}_{i \in \mathcal{N}}$
are the observation and action spaces, with $\mathcal{O}_i \subseteq \mathcal{S}, \; \forall i \in \mathcal{N}$. 
Further, $\left \{ \sigma_i \right \}_{i \in \mathcal{N}}$ 
and
$\left \{ \mathcal{R}_i \right \}_{i \in \mathcal{N}}$
are the agent observation and reward functions (potentially identical for all agents), such that
$\sigma_i : \mathcal{S} \mapsto \mathcal{O}_i$, and,
$\mathcal{R}_i: \mathcal{S} \times \left \{ \mathcal{A}_i \right \}_{i \in \mathcal{N}} \times \mathcal{S} \mapsto \R$.
$\mathcal{T}$ is the stochastic state transition model, defined as $\mathcal{T} : \mathcal{S} \times \left \{ \mathcal{A}_i \right \}_{i \in \mathcal{N}} \times \mathcal{S} \mapsto [0,1]$, which outputs the probability $\mathcal{T}(s^t, \left \{ a^t_i \right \}_{i \in \mathcal{N}},s^{t+1})$ of transitioning to state $s^{t+1}$ given the current state $s^t$ and actions $\left \{ a^t_i \right \}_{i \in \mathcal{N}}$. $\gamma$ is the discount factor.

Agents have a stochastic policy $\pi_{\theta_i}(a_i|o_i) $, which maps observations to action distributions that are sampled in the POMG to maximize the sum of discounted rewards received from the environment. The policy of agent $i$ is conditioned on neural network parameters $\theta_i$. To train policies $\pi_{\theta_i}$ we adopt the (Het)GPPO models presented in~\cite{bettini2023hetgppo}. Using GPPO, we perform homogeneous training via parameter sharing, thus $\theta = \theta_1 = \ldots = \theta_n$. Homogeneous agents are constrained to use the \textit{same} policy $\pi_\theta$ but benefit from the higher sample efficiency resulting from sharing parameters. On the other hand, when using HetGPPO, agents are able to learn independent heterogeneous policies $\pi_{\theta_i}$ and thus can develop behavioral differences. 
Note that, as mentioned before, homogeneous policies can sometimes \textit{emulate} diverse behavior by exploiting contextual information from their input~\cite{bettini2023hetgppo}, and, thus, our comparisons here will employ the same inputs for both homogeneous and heterogeneous models.

\subsection{Objective} 
The goal of this work is to measure the behavioral heterogeneity in a system of learning agents with heterogeneous policies $\pi_{\theta_i}$.
Our objective is to develop a metric,
$$
\mathrm{SND}: \left\{\pi_{\theta_i}\right\}_{i \in \mathcal{N}} \mapsto \R_{\geq 0},
$$
that takes as input the agents' policies and outputs a scalar value representing
the behavioral diversity of the system.
This metric will depend only on the learned agent policies and, thus, does not take the POMG components directly as input.
Since we are comparing policies of different agents, we assume that all agents have the same observation space $\mathcal{O} = \mathcal{O}_1=\ldots=\mathcal{O}_n$ and action space $\mathcal{A} =\mathcal{A}_1 = \ldots = \mathcal{A}_n$.
Importantly, differences in parameter space do not necessarily map to differences in behavioral space~\cite{stanley2019designing, kaplanis2019policy}, thus we cannot measure diversity directly through differences between parameters $\theta_i$ and $\theta_j$, but we need to measure it from policy outputs $\pi_{\theta_i}(o)$ and $\pi_{\theta_j}(o)$.
\section{Method: A Metric for System Neural Diversity}
Developing a metric to measure neural diversity in a multi-agent system is a challenging task. 
This is due to the fact that reducing a behavioral property dependent on potentially millions of neural connections to a single scalar value inevitably leads to high information loss.
Our goal is to minimize this information loss while retaining informative properties in the resulting metric.

In order to measure system diversity we first need a way to compare individuals. We thus tackle the following two tasks.
First, we define a measure of inter-agent pairwise \textbf{behavioral distance}.
We structure pairwise distances obtained with this measure in a behavioral distance matrix, which represents the distribution of agents' policies in behavioral space. 
Second, we aggregate the behavioral distance matrix into a single diversity value, which represents the \textbf{neural diversity metric} of the entire system.
In the following, we discuss the choice of the functions used for these two tasks.

\subsection{Behavioral Distance}

Heterogeneity is a collective concept. That is, it cannot be measured as an absolute property pertaining to a single agent in the system, but it has to be expressed as a relative measure among agents. 
We thus need to develop a mathematical metric that measures the behavioral distance between two agents $i,j \in \mathcal{N}$ as $d(i,j)$, where $d: \mathcal{N} \times \mathcal{N} \mapsto \R_{\geq 0}$. 
We want $d(i,j)$ to follow the properties of a mathematical metric~\cite{menger2003statistical}.

\begin{define}[Properties of the behavioral distance metric~\cite{menger2003statistical}]
\label{def:metric_properties}
For the distance $d$ to be a metric, it has to satisfy the following properties $\forall i,j,k \in \mathcal{N}$:
\begin{enumerate}
 \setlength\itemsep{0.1pt}
    \item  \textit{Non-negativity}: $d(i,j) \geq 0$ 
    \item  \textit{Identity of indiscernibles}: $d(i,j) = 0$ $\mathrm{iff}$ $\pi_{\theta_i}=\pi_{\theta_j} $ 
    \item  \textit{Symmetry}: $d(i,j) = d(j,i)$ 
    \item  \textit{Triangle inequality}: $d(i, j) \leq d(i, k) + d(k,j)$
\end{enumerate}
\end{define}

This set of properties aligns naturally with behavioral distance. We can think of two agents $i,j$ to be homogeneous when $d(i,j) = 0$ and increasingly diverse as the metric grows.


Our goal is to develop a distance
$$
d(i,j) = \int_\mathcal{O}f(\pi_{\theta_i}(o),\pi_{\theta_j}(o))\; do
$$
where a function $f$, providing a distance between two policies for a given observation, is evaluated over all possible observations.
Two key challenges follow:
(1) $f$ has to be a measure between the two probability distributions outputted by the polices and (2) evaluating distance over all possible continuous observations is intractable and, thus, we need a clever sampling strategy to create a subset of observations for evaluating the distance. In the following two subsections, we discuss how our proposed distance addresses these issues.

\subsubsection{Distance for Stochastic Policies}
Stochastic policies map a given observation to an action distribution.
Thus, a comparison of two agent behaviors can be made by measuring the statistical distance between their distributions.
There exist a variety of statistical distances.
Distances that follow properties 1-4 of \autoref{def:metric_properties} are referred to as \textit{metrics}, while statistical distances that only satisfy 1-2 are called \textit{divergences}. 
Divergences (such as the widespread Kullback–Leibler, KL) do not follow triangle inequality (property-4), which is useful when determining an upper bound of the behavioral distance between two agents whose respective distances to a third agent are known.
Among metrics, on the other hand, we focus our attention on the Wasserstein metric ($W_2$)~\cite{vaserstein1969markov}.
The $W_2$ distance between multivariate Gaussian action distributions can be computed in closed-form using the distributions' parameters. 
It represents the minimum cost required to move all the probability mass from one distribution to the other in an optimal transport problem formulation. 

\begin{figure}[t]
    \centering
    \includegraphics[width=0.6\columnwidth]{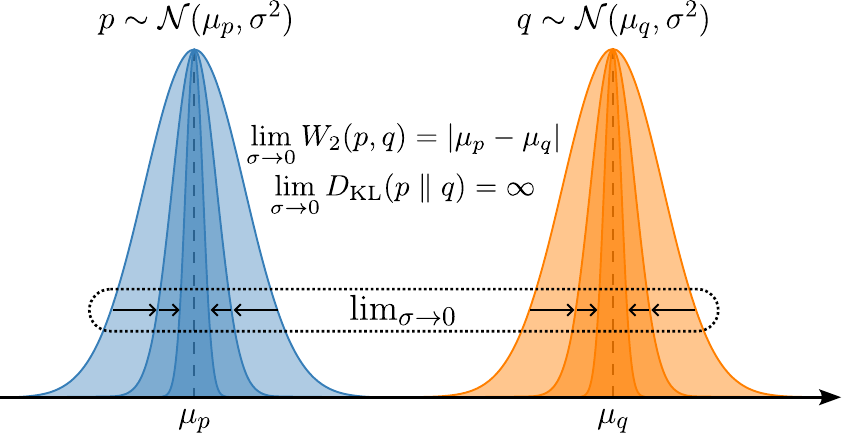}
    \caption{Wasserstein metric $W_2(p,q)$ and KL divergence $D_{\mathrm{KL}}(p\;\|\;q)$ of two univariate distributions $p,q$ as their standard deviation approaches 0. The value of $W_2$ in this scenario approaches the absolute difference of their means while KL divergence approaches infinity.}
    \label{fig:w_vs_kl}
\end{figure}

\autoref{fig:w_vs_kl} depicts an illustrative example to further elucidate the differences between the Wasserstein metric and KL divergence.
In this example, $p\sim\mathcal{N}(\mu_p,\sigma^2)$ and $q\sim\mathcal{N}(\mu_q,\sigma^2)$ are univariate Gaussian distributions. $W_2(p,q)$ represents the Wassertein metric, $D_{\mathrm{KL}}(p\;\|\;q)$ the KL divergence.
We observe that, as $\sigma$ approaches $0$, meaning that the probability mass of both distributions converges to their mean (as in Dirac delta distributions), $D_{\mathrm{KL}}(p\;\|\;q) \to \infty$ while $W_2(p,q) \to |\mu_p -\mu_q|$. This shows that, in the case the distributions assign increasing probability mass to their mean, a common scenario in MARL policies, Wasserstein outputs a bounded value proportional to the distance between the means, while KL does not. A similar argument was used to motivate Wasserstein generative adversarial networks~\cite{arjovsky2017wasserstein} where the model architecture significantly benefited from the use of $W_2$ over KL.

Following the reasons above, we use the Wasserstein metric to measure the probability distance between two policies\footnote{In our code we provide implementations of several metrics and distances, including the Hellinger distance~\cite{hellinger1909neue}, which can alternatively be used.}.
The resulting behavioral distance takes the following form:
\begin{equation}
\label{eq:d_integral_w}
    d(i,j) = \int_\mathcal{O} W_2(\pi_{\theta_i}(o),\pi_{\theta_j}(o))\; do.
\end{equation}


\subsubsection{Observation Sampling for Distance Evaluation}

Since the integral in \autoref{eq:d_integral_w} is theoretically computed over the entire continuous space of observations, we require a tractable sampling technique to generate a finite subset $\mathcal{B}\subseteq\mathcal{O}$ over which to evaluate the metric.
We choose to create $\mathcal{B}$ via rollouts (i.e., executions of the policy over time). In particular, every time we evaluate the behavioral distance, a set of rollouts is collected from the environment. A rollout is a collection of environment interactions stored in tuples of the form $\left \langle\{o_i^t\}_{i\in\mathcal{N}},\{a_i^{t}\}_{i\in\mathcal{N}},\{o_i^{t+1}\}_{i\in\mathcal{N}}\right \rangle$ with time $t \in [0,T]$.
We denote the sets of agents' observations and actions at time $t$ as $\mathbf{o}^t = \{o_i^t\}_{i\in\mathcal{N}}$ and $\mathbf{a}^t = \{a_i^t\}_{i\in\mathcal{N}}$, respectively. 
Thus, we construct $\mathcal{B} = \left\{\mathbf{o}^t \right\}_{t\in[0,T]}$.

When building $\mathcal{B}$, we want to avoid evaluating diversity on observations that were unseen by the agents during training. This is because their policies might present undefined behavior in such cases. 
Our process of sampling via environment rollouts in a Monte-Carlo fashion~\cite{sutton2018reinforcement} provides a high likelihood that these observations were seen previously during training.
While it has zero sampling bias, it is known to have a high variance in the states visited. We mitigate this by performing multiple rollouts. The number of rollouts performed has to be chosen based on the state distribution in the POMDP under evaluation.
Increasing the number of rollouts will decrease variance but increase the computational cost of evaluating the distance.
Now, we can write the final formulation of the behavioral distance $d$ as:

\begin{equation}
    d(i,j) = \frac{1}{|\mathcal{B}||\mathcal{N}|}\sum_{\mathbf{o}^t \in \mathcal{B}}\sum_{k \in \mathcal{N}} W_2(\pi_{\theta_i}(o_k^t),\pi_{\theta_j}(o_k^t)).
\end{equation}
This formulation states that the behavioral distance between agents $i$ and $j$ is the average Wasserstein metric computed between the distributions outputted by their policies over the observations of the agents collected over policy rollouts. 

\subsubsection{Behavioral Distance Matrix}
We can now structure these inter-agent distances in a behavioral distance matrix. Let
$$\mathbf{d}(i) = \left [ d(i,1), \ldots, d(i,n) \right ]\quad\quad n=|\mathcal{N}|$$
define the distances between $i$ and all other agents. We can then define the behavioral distance matrix as 
$$\mathbf{D} =\left [ \mathbf{d}(1)^\top, \ldots, \mathbf{d}(n)^\top \right ] .$$

Looking at the way this matrix is constructed, we can note some of its proprieties. Firstly, being constructed from a metric distance, it inherits the properties from \autoref{def:metric_properties}. In particular, this matrix is \textit{non-negative} (Property 1), \textit{hollow} (Property 2), and \textit{symmetric} (Property 3). Furthermore, by computing the sum of each row $\mathbf{d}(i)$, we can obtain a per-agent contribution to the system diversity. For example, in a system with $n$ agents where $d(1,j) = x$, for all $j \in\mathcal{N}\setminus \{1\}$ and $d(i,j) = 0$, for all $i,j \in\mathcal{N}\setminus \{1\}$, we get a contribution of $\sum_{j\in \mathcal{N}} d(1,j) = \frac{x(n-1)}{n}$ for agent $1$ and a contribution of $\sum_{j\in \mathcal{N}} d(i,j) = \frac{x}{n}$ for all other agents $i\in\mathcal{N}\setminus \{1\}$. By then calculating the relative fraction over the resulting values, we can express the percentage agent contributions to the team heterogeneity. 



\subsubsection{Example: Multi-Agent Goal Navigation}
\label{sec:navigation}

\begin{figure}[t]
    \centering
    \includegraphics[width=0.3\columnwidth,fbox]{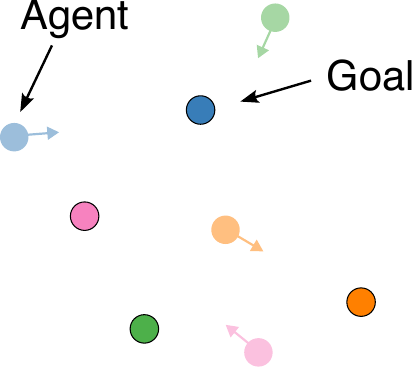}
    \caption{\textit{Multi-Agent Goal Navigation} example. Agents are spawned at random positions in a 2D workspace and take velocity actions (colored arrows) to reach their assigned goal, also spawned at random positions. }
    \label{fig:multi_goal}
\end{figure}

Let us now look at an experimental case study to complement the theoretical discussion on behavioral distance. 
In the \textit{Multi-Agent Goal Navigation} example, depicted in \autoref{fig:multi_goal}, $n$ agents are spawned at random positions in a 2D workspace. Each agent is assigned a goal, also spawned at random. Agents observe the relative position to all goals and output the mean and standard deviation of a 2D action distribution representing their desired velocity. This distribution is represented by two univariate Gaussians, which are sampled to get the action for each dimension. The reward for each agent is the difference in the relative distance to its goal over two consecutive timesteps, incentivizing agents to move towards their goals. 

\begin{figure*}[t]
    \centering
    \includegraphics[width=\columnwidth]{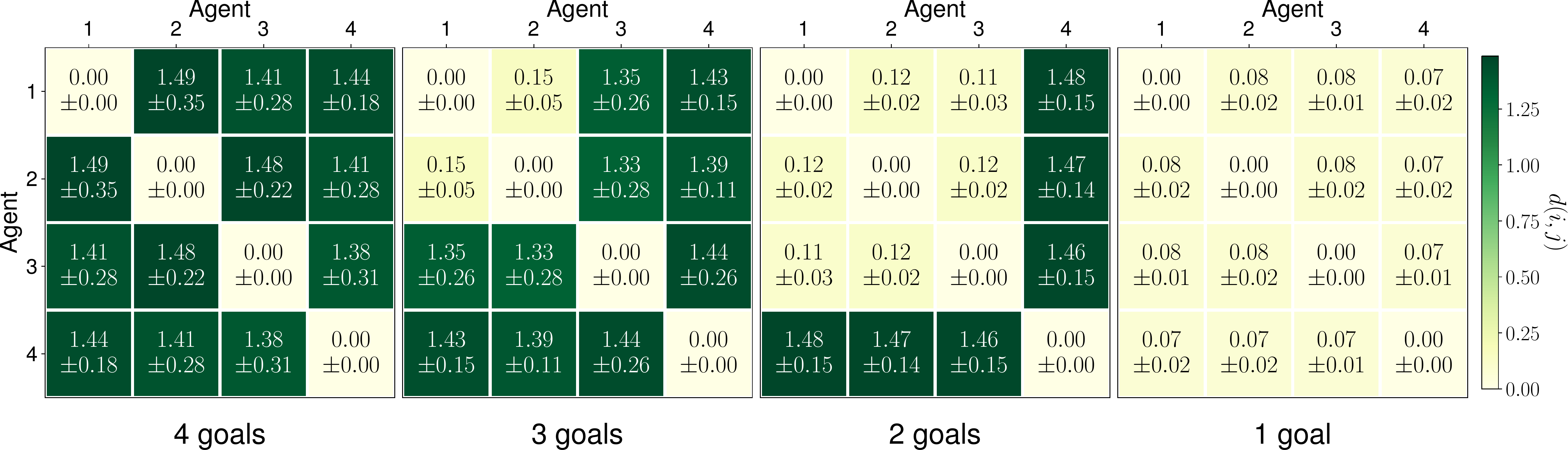}
    \caption{Behavioral distance matrices for the four experiments run on \textit{Multi-Agent Goal Navigation}. We can observe how, when agents are assigned the same goal, they become behaviorally homogeneous and thus reduce their behavioral distance. We report mean and standard deviation for $d(i,j)$ over 5 random seeds for each experiment. The values are collected after 300 training iterations each performed over 600 episodes of experience.}
    \label{fig:multi_goal_matrix}
\end{figure*}

We run four training experiments with $n=4$ and the following setups:
\begin{itemize}
 \setlength\itemsep{0pt}
    \item \textit{4 goals}: all agents are assigned a different goal
    \item \textit{3 goals}: agents $1,2$ are assigned the same goal and the rest have different goals
    \item \textit{2 goals}:  agents $1,2,3$ are assigned the same goal and the remaining agent has a different goal
    \item \textit{1 goal}: all agents are assigned the same goal
\end{itemize}
In \autoref{fig:multi_goal_matrix} we report the behavioral distance matrices for the four experiments. We can observe how, when agents are assigned the same goal, they learn the same policy, thus decreasing their behavioral distance.

\subsection{System Neural Diversity}
\label{subsec:SND}

In the following, we introduce System Neural Diversity (SND), a diversity metric that maps a behavioral distance matrix to a scalar diversity value.
SND takes inspiration from the Gini coefficient~\cite{gini1912variabilita} used in the field of economics. This coefficient provides a measure of statistical dispersion and was created to represent income inequality over a population. 
In a similar way, we represent diversity as behavioral dispersion using distances $d(i,j)$, allowing SND values to range form zero to infinity.
The proposed SND takes the form:
\begin{equation}
    \mathrm{SND}(\mathbf{D}) = \frac{2\sum_{i=1}^n\sum_{j=i+1}^n d(i,j)}{n(n-1)},
\end{equation}
where, due to the symmetry of $\mathbf{D}$ and the fact that its diagonal is zero, we can consider only the upper right triangle of the behavioral distance matrix during computation. SND can be interpreted as the mean behavioral distance over unique pairs of agents in the system.

We now introduce two key properties of SND which highlight its complementary nature with respect to HSE.

\begin{figure*}[t]
    \centering
    \begin{subfigure}{0.49\linewidth}
       \centering
       \includegraphics[width=\columnwidth]{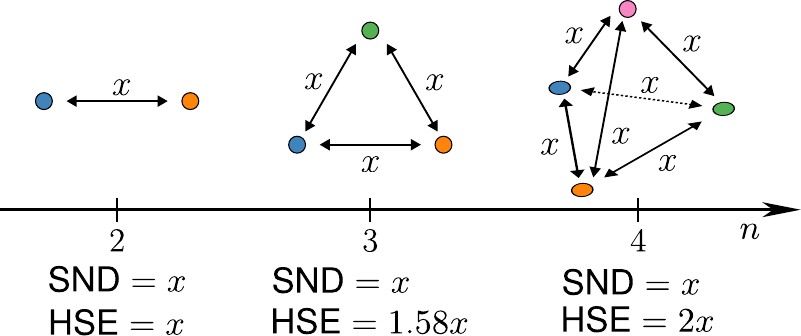}
       \caption{\textit{Invariance in the number of agents}: By modeling diversity as behavioral dispersion, SND does not increase as more agents ($n$, circles) are added at the same behavioral distance $x$ from each other. Conventional diversity metrics such as HSE~\cite{balch2000hierarchic} do not have this property.}
       \label{fig:invariance}
    \end{subfigure}\hfill
    \begin{subfigure}{0.49\linewidth}
        \centering
        \includegraphics[width=0.85\columnwidth]{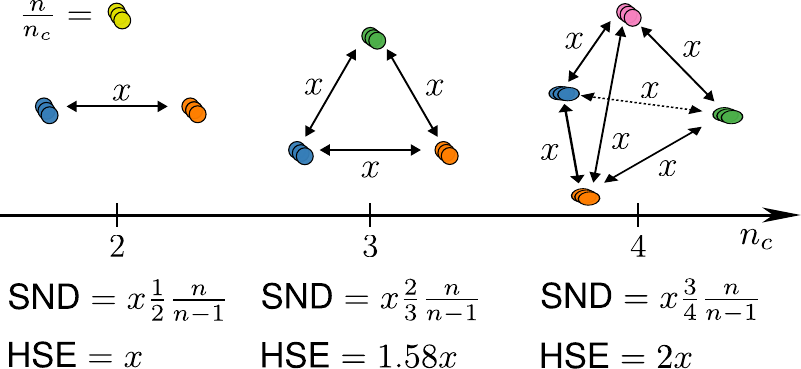}
        \caption{\textit{Redundancy measure}:
        Agents (circles) are divided into $n_c$ behavioral clusters (circle stacks) at distance $x$ in behavioral space for $n_c=2,3,4$. The reported values show that, for each value of $n_c$, SND decreases with $n$, while HSE is invariant to $n$. \autoref{fig:invariance} is a special case of this figure where $n=n_c$.}
        \label{fig:redundancy}
    \end{subfigure}
    \caption{An illustration of the properties of the proposed System Neural Diversity (SND) metric, contrasted against Heirarchical Social Entropy (HSE).}
    \label{fig:snd_properties}
\end{figure*}

\subsubsection{Invariance in the Number of Equidistant Agents}

When measuring heterogeneity, an important question is: ``\textit{If in a system with two agents at behavioral distance $x$ we add a third agent, also at distance $x$ from the other two, does the system's heterogeneity increase?}''. While HSE answers positively, SND provides a negative answer. This is because, aligning with the economic interpretation of the Gini coefficient, SND considers diversity as the behavioral dispersion of the system. Maximum dispersion is independent from the number of individuals considered. We refer to this property as \textit{invariance in the number of equidistant agents}.

\begin{property}[Invariance in the number of equidistant agents \autoref{fig:invariance}]
\label{prop:invariance}
Given a behavioral distance matrix $\mathbf{D}$, where $d(i,j) = x$, $\forall i,j \in \mathcal{N} $ with $i \neq j$, representing a system with all agents at behavioral distance $x$ from each other, $\mathrm{SND}(\mathbf{D})$ is invariant with respect to the number of agents $n$ in the system.
\end{property}
\begin{proof}
\label{app:proof_invariance}
Given that  $d(i,j) = x$, $\forall i,j \in \mathcal{N} $  with $i \neq j$, we can write 
$ \sum_{i=1}^n\sum_{j=i+1}^n d(i,j) = x \frac{n(n-1)}{2} $. Substituting in SND we get $\mathrm{SND}(\mathbf{D}) = \frac{2n(n-1)x}{2n(n-1)} = x$, which is not dependent on $n$.
\end{proof}

\autoref{fig:invariance} depicts this property by showing the SND and HSE values for $n=2,3,4$. The desirability of this property may depend on the use case. Nevertheless, it highlights how SND provides a complementary and additional tool to HSE when measuring diversity.

\subsubsection{A Measure of Behavioral Redundancy}
In heterogeneous systems, multiple agents might specialize in the same core skill in order to provide redundancy. For example, if we consider a population survival scenario, some agents may be required to forage food, while others may need to stay static and monitor an area. In games like football, agents may distribute between defenders and attackers. 
While it seems intuitive that behavioral redundancy should lead to a decrease in system diversity, this aspect is not captured in HSE. HSE clusters behaviorally identical agents together and then computes the Shannon entropy over the population distribution in the behavioral clusters. Thus, if $n_c$ behavioral clusters are present, with $\frac{n}{n_c}$ agents at distance 0 from each other in each of them, HSE will not depend on $n$, and thus not measure behavioral redundancy. On the other hand, SND is able to measure redundancy.

\begin{property}[Redundancy measure \autoref{fig:redundancy}]
\label{prop:redundancy}
Given a behavioral distance matrix $\mathbf{D}$, where $n$ agents are divided equally in $n_c$ behavioral clusters $\mathcal{C} = \left \{ \mathcal{C}_1, \ldots, \mathcal{C}_{n_c} \right \}$ with $|\mathcal{C}_h| = \frac{n}{n_c} \in \mathbb{N}_{> 0}$, $\forall \mathcal{C}_h \in \mathcal{C}$, $c(i): \mathcal{N} \mapsto \mathcal{C}$ is a function mapping each agent to its cluster, and
\begin{align*}
    d(i,j) = 
\left\{\begin{matrix} 
 0 & \quad\text{if } c(i) = c(j)
\\   x& \quad\text{otherwise}
\end{matrix}\right. ,
\end{align*}
SND is a monotonically decreasing function of $n$ and a monotonically increasing function of $n_c$, and it takes the form

$$
\mathrm{SND}(\mathbf{D}) = x\frac{n(n_c-1)}{n_c(n-1)}.
$$

\end{property}
\begin{proof}
\label{app:proof_redundancy}
Given that $n$ agents are equally distributed in $n_c$ behaviorally equidistant clusters, $ \sum_{i=1}^n\sum_{j=i+1}^n d(i,j) $ can be rewritten as $x  \frac{n^2}{n_c^2} \frac{n_c (n_c -1)}{2}$. Meaning that each pair of agents from two different clusters ($\frac{n^2}{n_c^2} $) is at distance $x$  for each unique pair of clusters ($\frac{n_c (n_c -1)}{2}$). Simplifying, we get  $ \sum_{i=1}^n\sum_{j=i+1}^n d(i,j) = \frac{xn^2(n_c-1)}{2n_c} $. Substituting in SND we get $\mathrm{SND}(\mathbf{D}) = \frac{2xn^2(n_c-1)}{2n_cn(n-1)} = x\frac{n(n_c-1)}{n_c(n-1)}$, where $\frac{n}{n-1}$ is monotonically decreasing function of $n$ and $\frac{n_c -1}{n_c}$ is a monotonically increasing function of $n_c$.
\end{proof}

In other words, SND increases with the number of behavioral clusters, and decreases with the number of agents per cluster.
\autoref{fig:redundancy} depicts this property by showing the SND and HSE as a function of $n$ for $n_c=2,3,4$.

\subsubsection{Example: Multi-Agent Goal Navigation}
\label{sec:navigation2}

To exemplify empirically the system-level view provided by SND, we consider again the \textit{Multi-Agent Goal Navigation} scenario from \autoref{sec:navigation}. We utilise the same setup with $n=4$ agents and an increasing number of goals.

\begin{figure}[t]
    \centering
    \begin{subfigure}{0.4\linewidth}
       \includegraphics[width=\columnwidth]{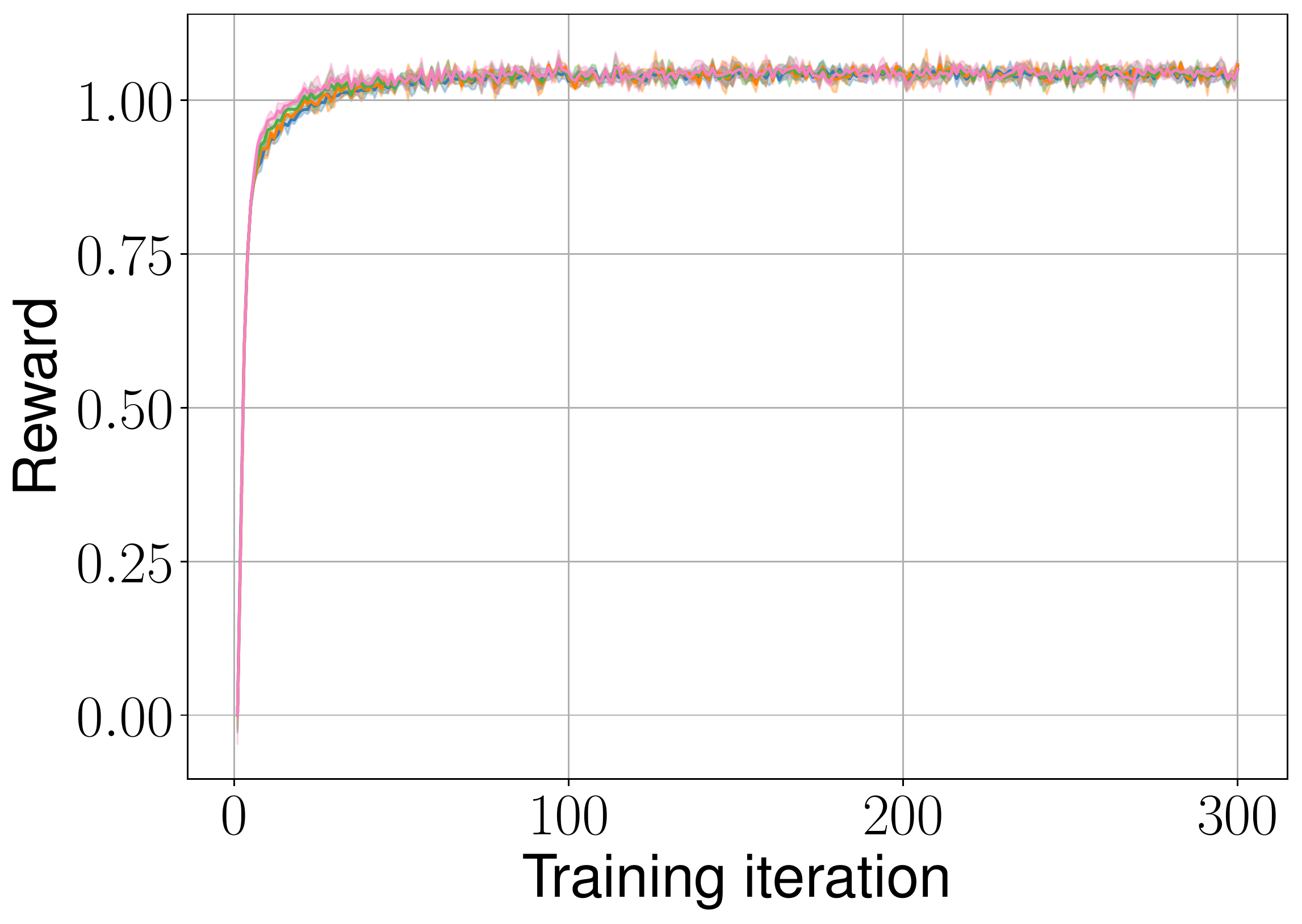}
       \caption{Reward}
          \label{fig:static_multi_goal_rew}
    \end{subfigure}
    \begin{subfigure}{0.4\linewidth}
        \includegraphics[width=\columnwidth]{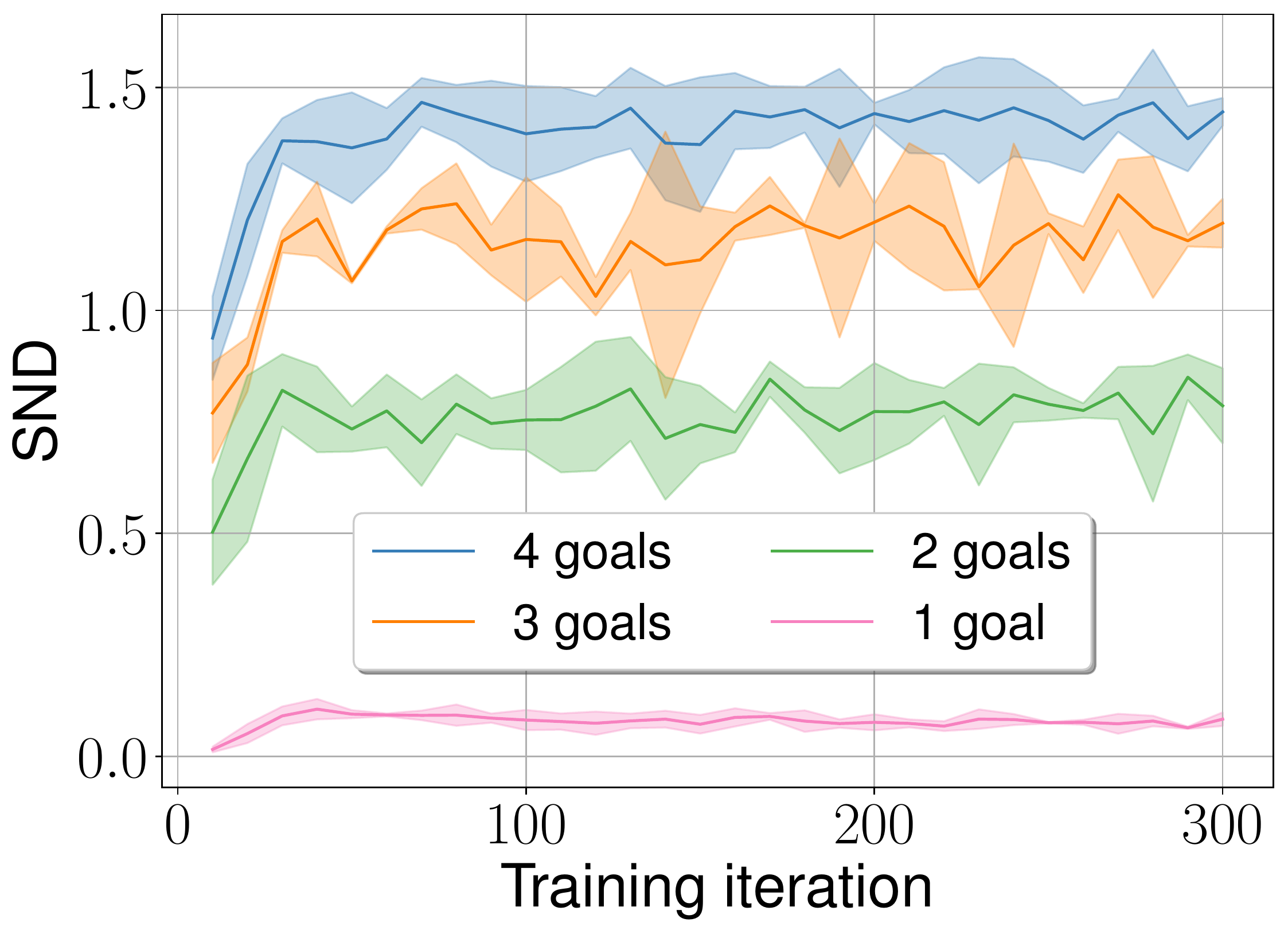}
        \caption{System Neural Diversity}
        \label{fig:static_multi_goal_snd}
    \end{subfigure}\hfill
    
    \caption{SND in the \textit{Multi-Agent Goal Navigation} scenario. We can observe that, while all setups reach the same reward, SND decreases as the agents share more goals, until the system becomes homogeneous when all agents are sharing the same goal.
    We report mean and standard deviation over 5 random seeds for each experiment. The values are measured over 300 training iterations each performed over 600 episodes of experience.}

\end{figure}

The SND measured throughout training (\autoref{fig:static_multi_goal_snd}) decreases with the number of goals.
This means that, as the number of goals decreases, more agents will be assigned the same goal and will thus develop more homogeneous navigation policies.
In particular, when the agents all share one goal, SND approaches 0.
An SND value of 0 indicates that the agents are behaving homogeneously.
In such a case, the metric acts an important diagnostic tool, suggesting that a homogeneous training strategy should be preferred in subsequent experiment iterations in order to benefit from parameter sharing and increased sample efficiency. More details on the sample efficiency benefits of homogeneous training are reported in \autoref{app:hom_train}.

\section{Results}

We are now interested in using SND as a heterogeneity metric to measure, control, and develop insights on team diversity during a MARL training phase.
For this, we run a series of experiments on various multi-robot tasks. These generally represent multi-robot coordination problems with POMGs that require inter-agent communication to be solved.
We analyze two types of tasks: (1) \textit{static tasks} where the agents have to solve a problem that does not change throughout training, and
(2) \textit{dynamic tasks} where the problem can change throughout training due to unmodeled disruptions (i.e. noise, external forces, adversaries, etc.).
Agents are trained using HetGPPO~\cite{bettini2023hetgppo}, HetIPPO (HetGPPO without communication) and their homogeneous counterparts: IPPO and GPPO. The simulation environments representing these tasks are partly new implementations and partly adapted from existing environments in the VMAS benchmark set~\cite{bettini2022vmas}.

\subsection{Comparison to HSE}

We begin by contrasting SND and HSE in the \textit{Multi-Agent Goal Navigation} task (\autoref{fig:multi_goal}).
These provide empirical demonstrations and motivation for the theoretical properties of SND (\autoref{subsec:SND}).

\subsubsection{Invariance in the number of equidistant agents}

To showcase \textit{Invariance in the number of equidistant agents} experimentally, we implement HSE (using our $W_2$ as the behavioral distance  to account for stochastic policies) and run 7 experiments in the \textit{Multi-Agent Goal Navigation} (\autoref{sec:navigation}) with an increasing number of agents $n$, each with one goal.
\autoref{tab:invariance} shows that SND value remains invariant to the increasing addition of new agents with new goals, while HSE grows.

\begin{table}[bp]
\centering
\caption{Invariance in the number of equidistant agents in the \textit{Multi-Agent Goal Navigation} scenario. We experimentally show that SND is invariant to the increasing addition of new agents to the system, while HSE grows. We report mean and standard deviation over 4 random seeds for each $n$. The values are collected after 300 training iterations each performed over 600 episodes of experience.}
\label{tab:invariance}
\begin{tabular}{lccccccc}
$n$ & 2 & 3 & 4 & 5 & 6 & 7 & 8 \\
\toprule
SND &
\parbox{0.07\textwidth}{\centering$1.51$\\$\pm0.14$} &
\parbox{0.07\textwidth}{\centering$1.46$\\$\pm0.11$} &
\parbox{0.07\textwidth}{\centering$1.43$\\$\pm0.07$} &
\parbox{0.07\textwidth}{\centering$1.45$\\$\pm0.06$ }&
\parbox{0.07\textwidth}{\centering$1.42$\\$\pm0.06$ }&
\parbox{0.07\textwidth}{\centering$1.44$\\$\pm0.06$ }&
\parbox{0.07\textwidth}{\centering$1.43$\\$\pm0.05$ }\\\midrule
HSE &
\parbox{0.07\textwidth}{\centering$1.51$\\$\pm0.14$ }&
\parbox{0.07\textwidth}{\centering$2.50$\\$\pm0.18$ }&
\parbox{0.07\textwidth}{\centering$3.17$\\$\pm0.23$ }&
\parbox{0.07\textwidth}{\centering$3.86$\\$\pm0.21$ }&
\parbox{0.07\textwidth}{\centering$4.32$\\$\pm0.24$ }&
\parbox{0.07\textwidth}{\centering$4.77$\\$\pm0.31$ }&
\parbox{0.07\textwidth}{\centering$5.21$\\$\pm0.29$ } \\
\bottomrule
\end{tabular}
\end{table}

\subsubsection{A Measure of Behavioral Redundancy}

\begin{table}[ht]
\centering
\caption{Redundancy measure in the \textit{Multi-Agent Goal Navigation} scenario with $n_c=2$. 
We experimentally show that while SND decreases with the redundancy of agents in clusters, HSE slightly increases. We report mean and standard deviation over 6 random seeds for each $n$. The values are collected after 300 training iterations each performed over 600 episodes of experience.}
\label{tab:redundancy}
\begin{tabular}{lcccc}
$n$ & 2 & 4 & 6 & 8  \\
\midrule
SND & $1.49\pm0.12$ & $0.98\pm0.09$ & $0.85\pm0.07$ & $0.81\pm0.06$ \\ \midrule
HSE & $1.49\pm0.12$ & $1.65\pm0.17$ & $1.76\pm0.16$ & $1.91\pm0.16$ \\
\bottomrule
\end{tabular}

\end{table}

To showcase how SND measures behavioral redundancy experimentally, we modify the \textit{Multi-Agent Goal Navigation} scenario (\autoref{sec:navigation}) by fixing the number of goals to 2, corresponding to two behavioral clusters ($n_c=2$). We then run experiments with $n=\{2,4,6,8\}$ in which the first half of the team is assigned to the first goal and the second half to the other.
\autoref{tab:redundancy} shows that SND decreases with the number of agents,
thus capturing a redundancy in behavioral diversity as more agents behave similarly.
On the other hand, HSE not only does not decrease, but slightly increases, as agents with the same goal will be at behavioral distance $\approx0$.

\subsection{Static Tasks}
Next, we train agents on a set of static tasks that benefit from behavioral diversity, and use SND to analyze the impact of such diversity in the learning process.
We use the term \textit{static tasks} to refer to multi-agent problems modeled by a fixed\footnote{A POMG that does not vary throughout the learning process.} POMG. 

\begin{figure*}[h!]
    \centering
    \begin{subfigure}[b]{0.25\linewidth}
        \centering
        \includegraphics[width=0.6\linewidth]{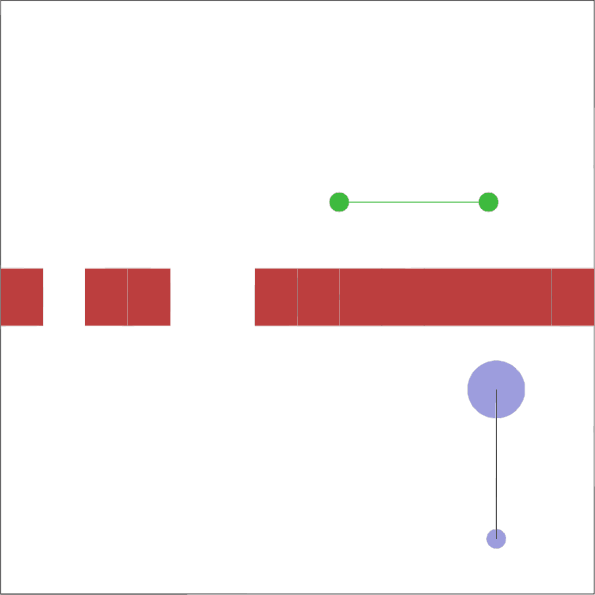}
        \vspace*{2.5ex}
        \caption{Different Size Joint Passage}
        \label{fig:static_size}
    \end{subfigure}%
    \begin{subfigure}[b]{0.35\linewidth}
        \centering
       \includegraphics[width=0.9\columnwidth]{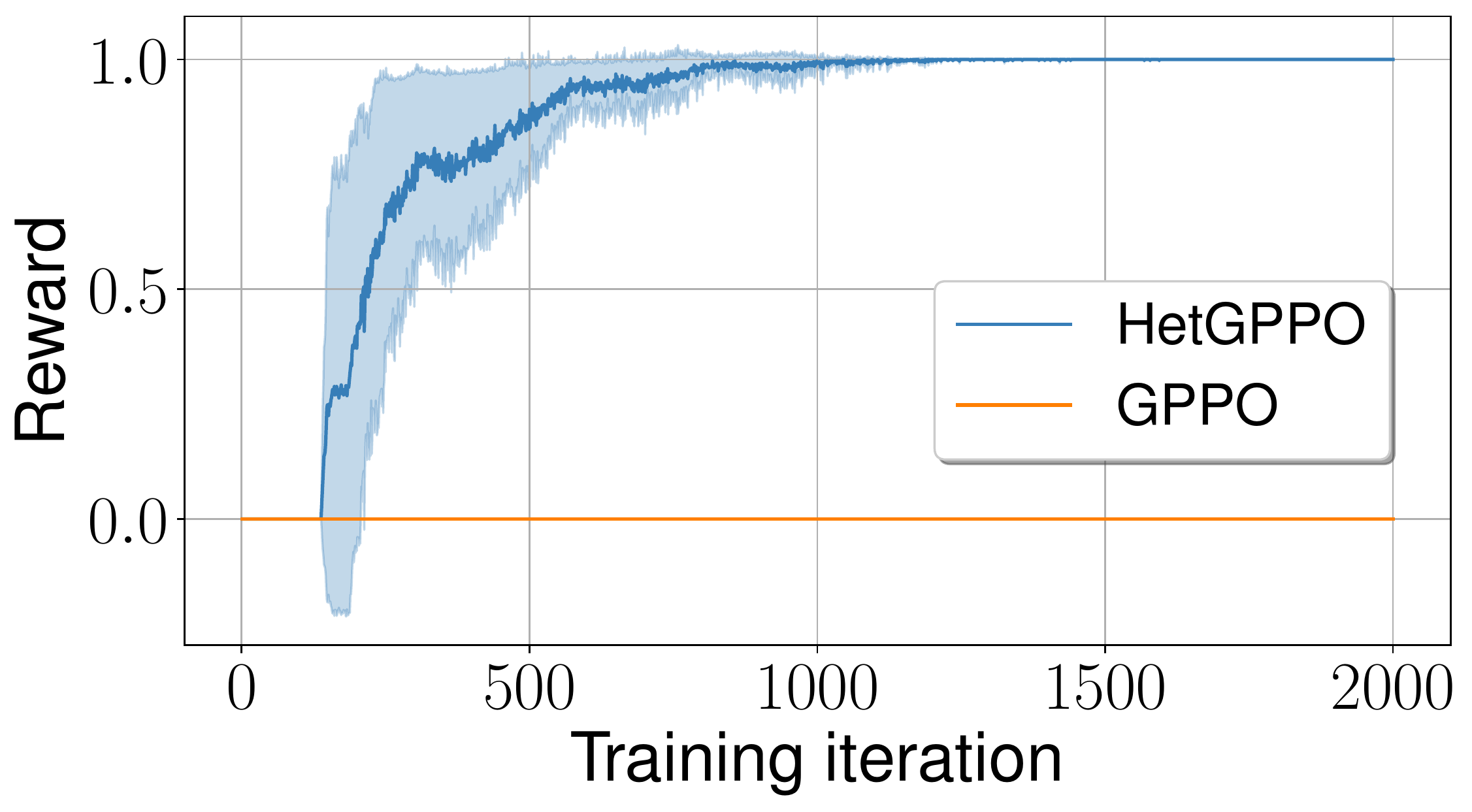}
       \caption{Training Reward.}
        \label{fig:static_size_reward}
    \end{subfigure}
    \begin{subfigure}[b]{0.35\linewidth}
        \centering
       \includegraphics[width=0.9\columnwidth]{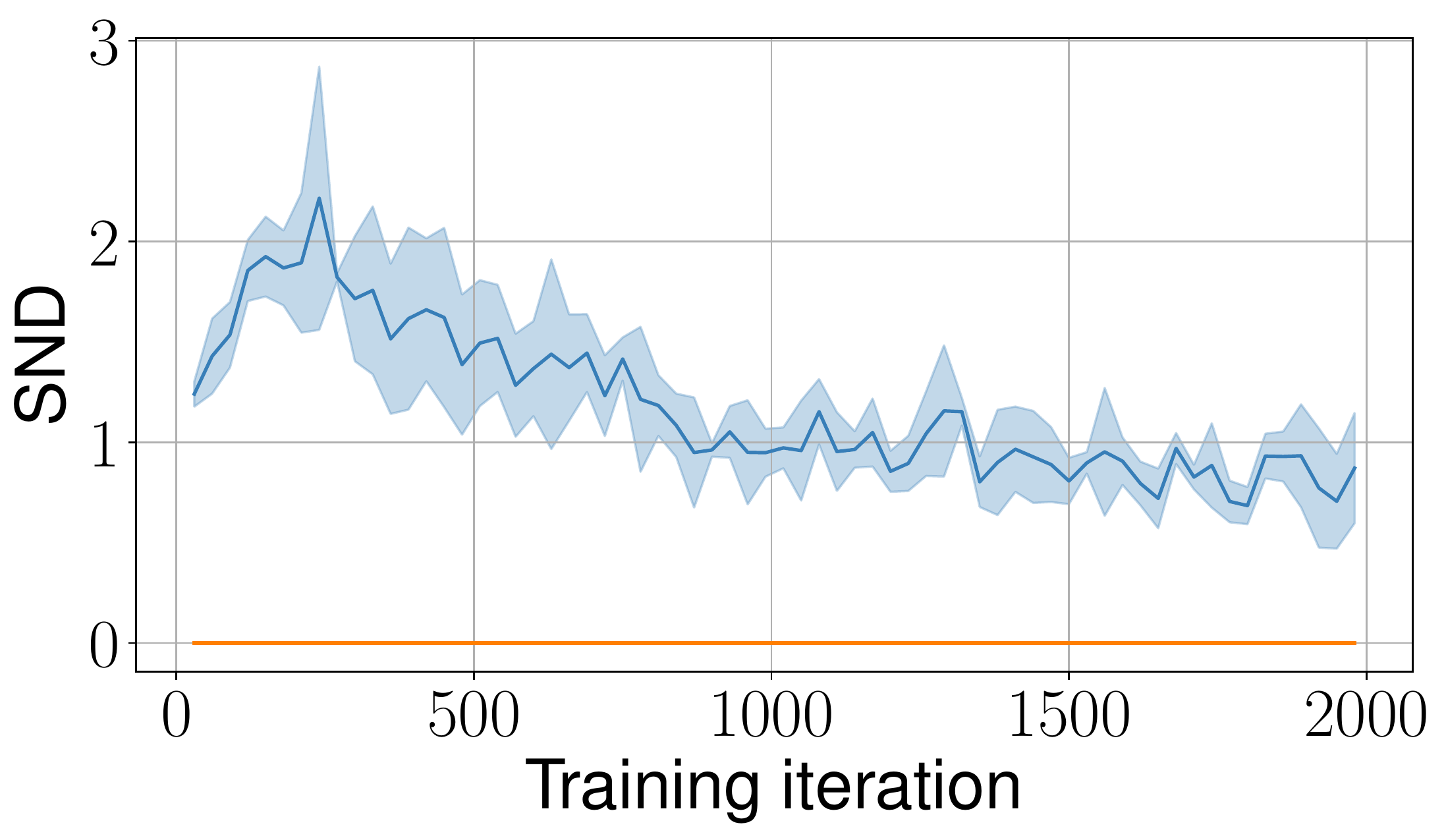}
       \caption{Diversity Metric.}
        \label{fig:static_size_snd}
    \end{subfigure}

    \centering
    \begin{subfigure}[b]{0.25\linewidth}
         \centering
        \includegraphics[width=0.6\linewidth]{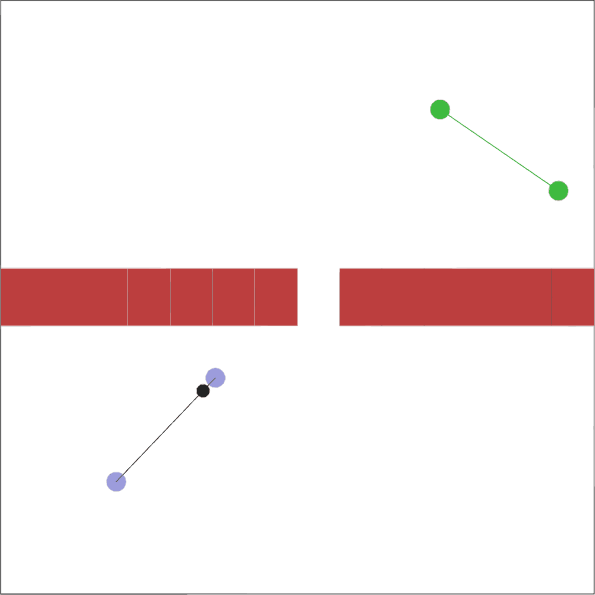}
        \vspace*{2.5ex}
        \caption{Asym. Payload Joint Passage}
         \label{fig:static_payload}
    \end{subfigure}
    \begin{subfigure}[b]{0.35\linewidth}
         \centering
       \includegraphics[width=0.9\columnwidth]{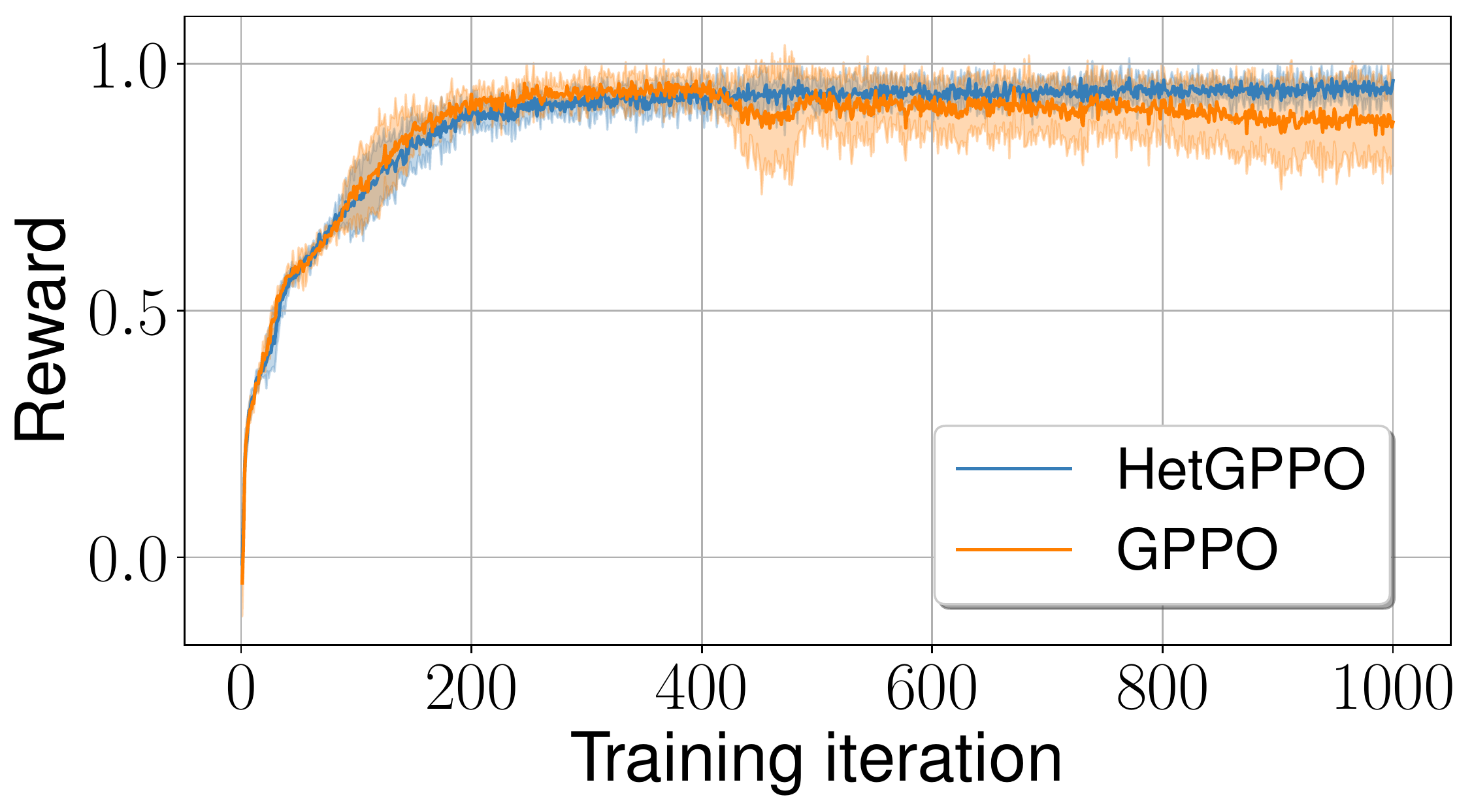}
       \caption{Training Reward.}
        \label{fig:static_payload_reward}
    \end{subfigure}
    \begin{subfigure}[b]{0.35\linewidth}
         \centering
       \includegraphics[width=0.9\columnwidth]{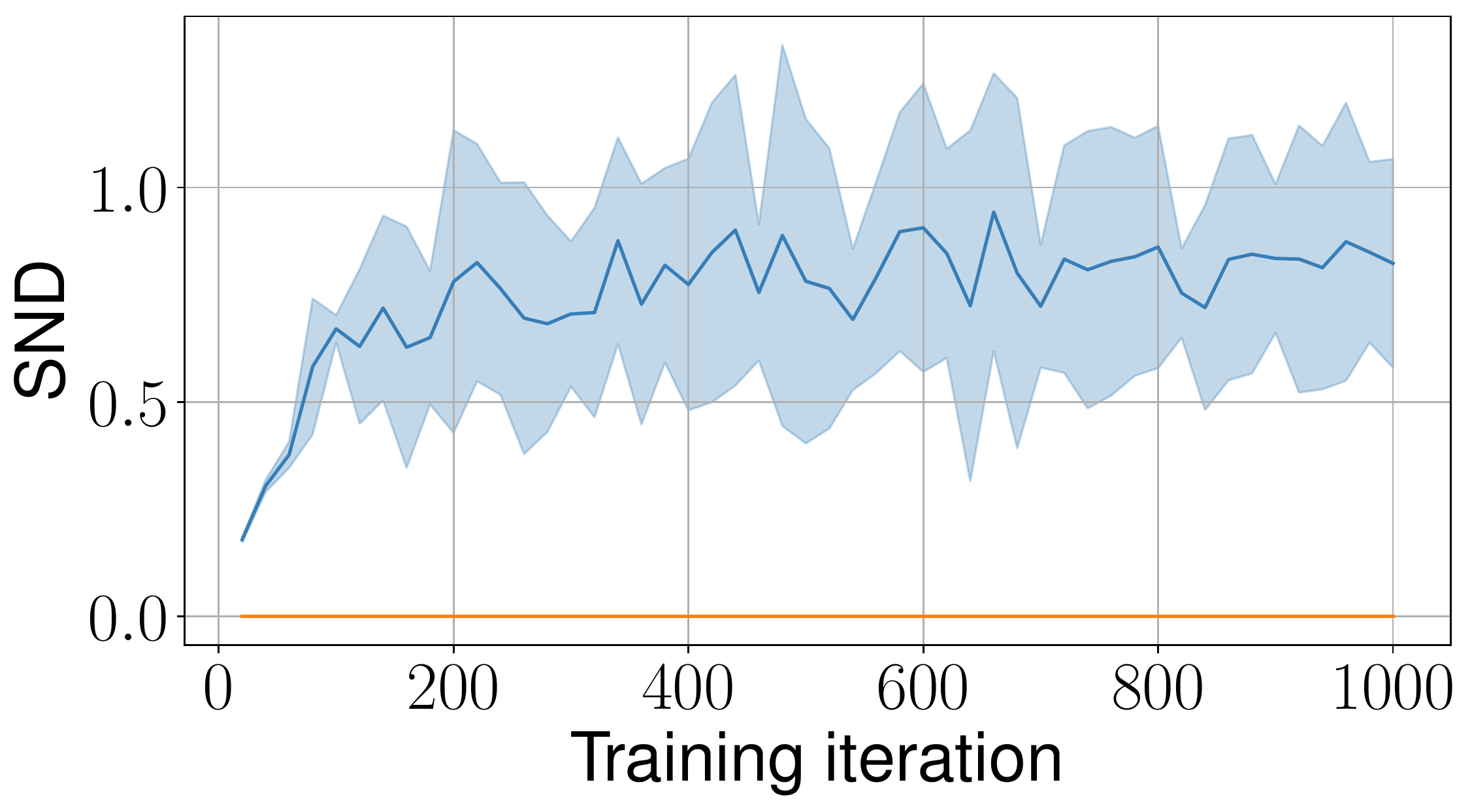}
       \caption{Diversity Metric.}
        \label{fig:static_payload_snd}
    \end{subfigure}
    \caption{Evaluations on static tasks. \textbf{Top row (left to right)}: \textit{Different Size Joint Passage} scenario with reward and SND. \textbf{Bottom row (left to right)}: \textit{Asymmetric Payload Joint Passage} scenario with reward and SND. These evaluations in static scenarios show that heterogeneity can grant performance improvements (top) and resilience improvements (bottom paired with \autoref{fig:asym_mass_noise_inject}).
    We report mean and standard deviation over 3 random seeds for each experiment. Each training iteration is performed over 600 episodes of experience.}
    \label{fig:static_tasks}
\end{figure*}

\subsubsection{Different Size Joint Passage}
\label{subsec:different_size_passage}
This task, shown in \autoref{fig:static_size}, involves two robots of different sizes (blue circles), connected by a rigid linkage through two revolute joints.
The team needs to cross a passage while keeping the linkage parallel to it and then match the desired goal position (green circles) on the other side.
The passage is comprised of a bigger and a smaller gap, which are spawned in a random position and order on the wall, but always at the same distance between each other.
The team is spawned in a random order and position on the lower side with the linkage always perpendicular to the passage.
The goal is spawned horizontally in a random position on the upper side. Each agent observes and communicates its velocity, relative position to each gap, and relative position to the goal center. The relative positions and velocities to the other agents are obtained through communication. The sizes of the agents or of the gaps are not part of the observations.
The reward function is global and shared by the team. It is composed of two convex terms: before the passage, the robots are rewarded to keep the linkage parallel to the goal and to carry its center to the center of the passage; after the passage, the robots are rewarded for carrying it to the goal at the desired orientation. Collisions are also penalized.

In \autoref{fig:static_size_reward} we show the training reward (proportional to the percentage of episodes in each batch that complete the task).
The plotted reward shows that this task requires heterogeneous behavior to be solved. In fact, the homogeneous agents, unable to observe their physical differences, cannot learn specialized roles, which would allow them to assign the smaller agent to the smaller gap. Agents with homogeneous policies never manage to cross the passage, being deterred by unavoidable collisions.
With the heterogeneous model, on the other hand, each agent is able to learn a specialized role and tackle the respective gap.
This is confirmed by observing SND in \autoref{fig:static_size_snd}, which indicates that the agents behave heterogeneously,
but learn to smooth out their diversity through training. This is due to the fact that diversity is needed only during the initial `assignment' action sequence (to position themselves with respect to the correct gap) and the rest of the navigation task can be solved homogeneously.

\subsubsection{Asymmetric Payload Joint Passage}
\label{subsec:asymm_payload_passage}
In the previous task, agents were physically different. In this task, we run an evaluation in a scenario where agents are physically identical but are impacted in diverse ways by the environment.
We consider the task depicted in \autoref{fig:static_payload}.
The setup of this task is similar to the \textit{Different Size Joint Passage} scenario, with the difference that the robots are now physically identical, but the linkage has an asymmetrically positioned payload (black circle).
The passage now is a single gap, located randomly in the wall.
The agents need to cross it while keeping the linkage perpendicular to the wall and avoiding collisions.
The team and the goal are spawned in a random position, order, and rotation on opposite sides of the passage.
Each robot observes and communicates its velocity, relative position to the gap, relative position to the goal center and goal orientation.
The relative positions and velocities to the other agents are obtained through communication.
The reward is shared and global and composed of two convex terms: before the passage, the team is rewarded for keeping the linkage perpendicular to the wall and moving towards the center of the gap.
After the passage, the team is rewarded for aligning with the goal position and orientation. 

\begin{figure}[t]
    \centering
    
    \begin{subfigure}{0.3\linewidth}
      \includegraphics[width=\linewidth]{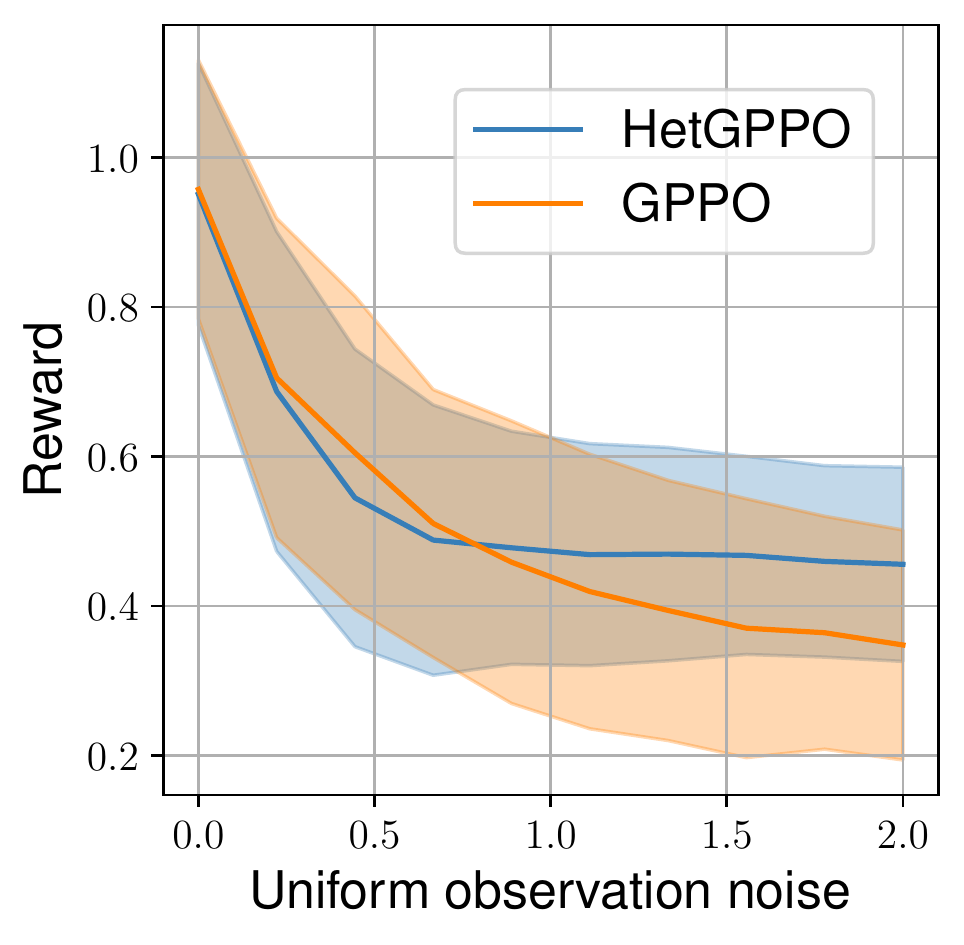}
       \caption{Reward degradation}
    \end{subfigure}
    \begin{subfigure}{0.3\linewidth}
        \includegraphics[width=\columnwidth]{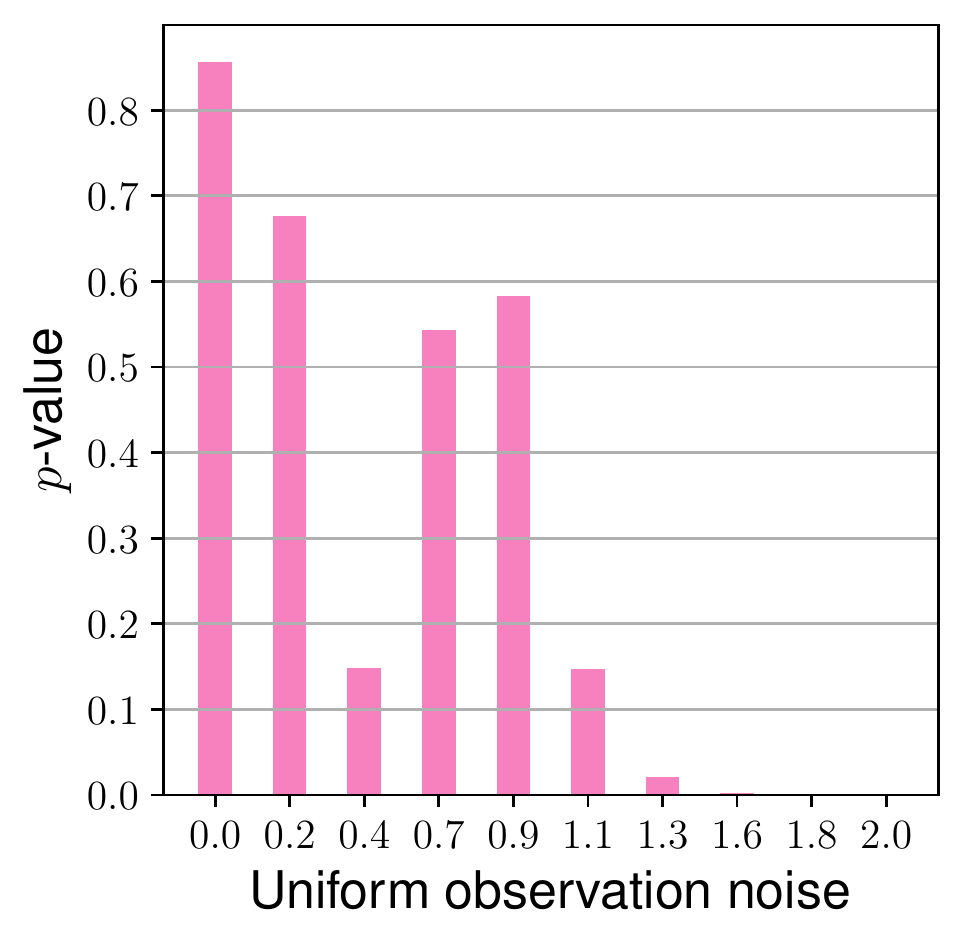}
        \caption{Welch's $t$-test}
    
    \end{subfigure}\hfill
    \caption{Performance degradation in the \textit{Asymmetric Payload Joint Passage} scenario in \autoref{fig:static_payload} in the presence of deployment noise.  We apply uniform observation noise $\mathcal{U}(-\delta,\delta)$ in the same units as the observations for 10 values of $\delta$ in the range $[0,2]$. For each $\delta$ we report the mean and standard deviation of the reward for 50 episodes (left) and perform a Welch's unequal variances $t$-test~\cite{welch1947generalization} for the means of the samples collected with the two models (right).} 
    \label{fig:asym_mass_noise_inject}
\end{figure}

By looking at the reward curve for this scenario in \autoref{fig:static_payload_reward}, we can observe that both heterogeneous and homogeneous agents are able to solve the task and obtain the maximum reward.
This is because the homogeneous model is able to infer the agent differences from physical observations through a process called \textit{inferred behavioral typing}~\cite{bettini2023hetgppo} and thus learn a single multi-behavioral policy conditioned on these differences.
In the heterogeneous model, on the other hand, two diverse policies are learned, leading to a significant difference in SND (\autoref{fig:static_payload_snd}).
This leads to the following question: ``\textit{Given that the two models achieve the same performance with different diversity scores, what advantages, if any, does heterogeneity offer in this scenario?}''.
To answer this question, we take the frozen learned policies for both models and evaluate them under increasing deployment noise.
In other words, we inject uniform observation noise at test time.
The result, reported in \autoref{fig:asym_mass_noise_inject}, shows that the heterogeneous model proves more resilient to increasing noise\footnote{Similar resilience results have been observed in~\cite{bettini2023hetgppo} in various scenarios.}.
From the $p$-values of a Welch's unequal variances $t$-test~\cite{welch1947generalization} we can observe that the performance curves are statistically different.
The low $p$-values for high noise injections suggest that we fail to accept the null hypothesis of the samples having the same mean.

\begin{figure*}[!h]
    \centering
    \begin{subfigure}{0.25\linewidth}
    \centering
        \includegraphics[width=0.65\linewidth]{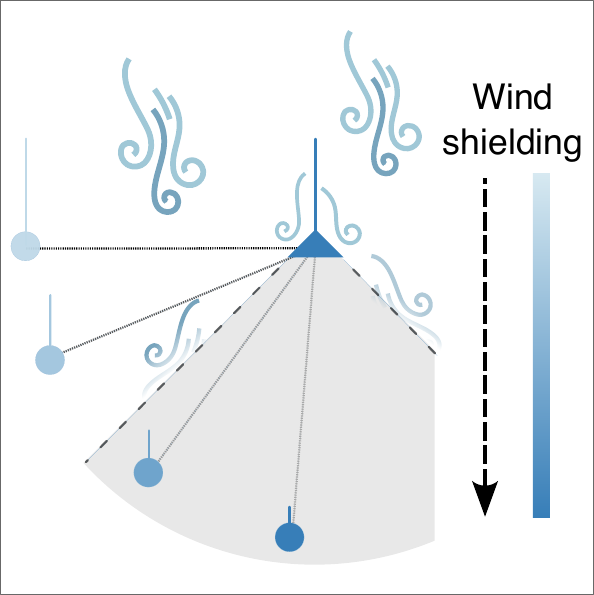}
        \vspace*{3ex}
        \caption{Flocking in Wind}
        \label{fig:wind_scenario}
    \end{subfigure}\hfill
    \begin{subfigure}{0.36\linewidth}
    \centering
       \includegraphics[width=\columnwidth]{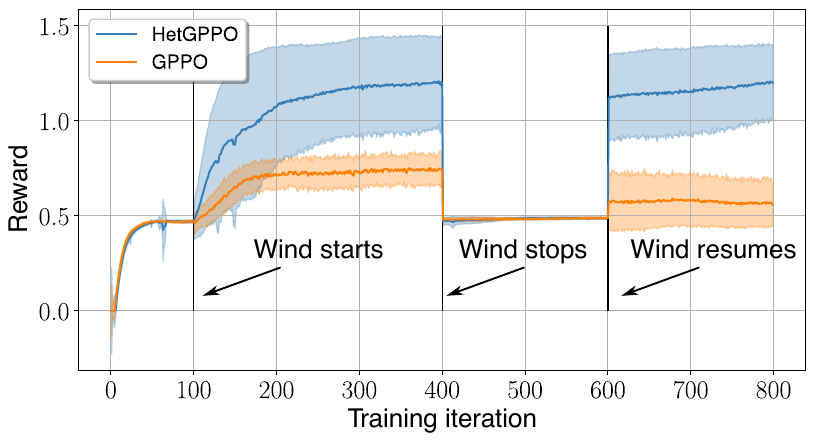}
       \caption{Reward}
       \label{fig:wind_reward}
    \end{subfigure}\hfill
    \begin{subfigure}{0.36\linewidth}
    \centering
       \includegraphics[width=\columnwidth]{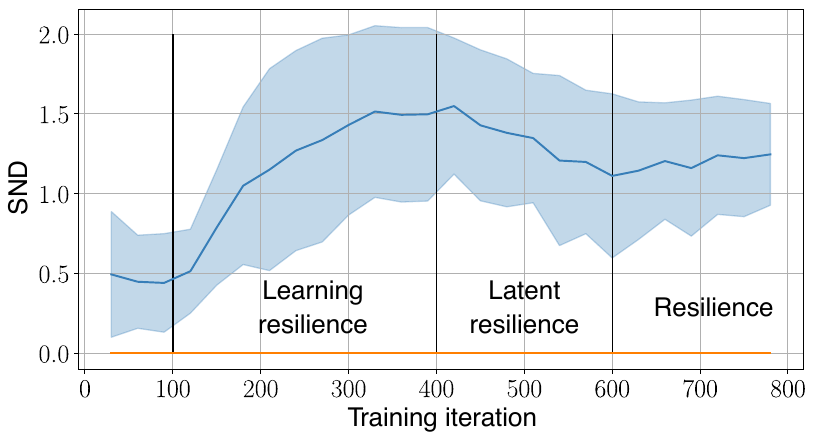}
       \caption{System Neural Diversity}
        \label{fig:wind_snd}
    \end{subfigure}
   
    \caption{Evaluations on the \textit{Flocking in Wind} dynamic task. These plots show the emergence of \textit{latent resilience} for heterogeneous learning. Heterogeneous agents are able to acquire resilience skills when facing a disturbance and utilize those skills in case the disturbance reappears.
    We report mean and standard deviation over 11 random seeds for each experiment. Each training iteration is performed over 300 (wind) episodes of experience.}
 
\end{figure*}

\subsection{Dynamic Tasks: Flocking in Wind}
\label{subsec:wind}
We use the term \textit{dynamic tasks} to refer to multi-agent problems modeled by a dynamic POMG (i.e., a POMG that can vary throughout execution).
Modifications to the POMG could occur due to unmodeled external disturbances such as noise, faults, or adversaries.
These modifications could disrupt the multi-agent system, which would then need to undergo an adaptation process (e.g., changing formation, communication topology, etc.) to regain performance.
We refer to the property of adaptation in the face of a disruption as \textit{resilience}~\cite{prorok2021beyond}.
Heterogeneous training in MARL has previously been shown to have resilience properties in numerous simulated and real-world scenarios~\cite{bettini2023hetgppo}. 
In this section, we are interested in analyzing the adaptation of a multi-agent system to a disruption during the training process.
In particular, we consider scenarios were such adaptation requires heterogeneous behaviors, and study how resilience relates to SND.

In the \textit{Flocking in Wind} scenario, shown in \autoref{fig:wind_scenario}, we consider a flock of $n=2$ agents (blue circle and blue triangle) in 2D space tasked with tracking a desired velocity vector of \SI{0.5}{m/s} directed North (top of the figure) while keeping a desired distance of \SI{1}{m} (dotted line between the agents).
The agents receive a shared global reward proportional to the reduction in the errors from the reference velocity and team distance every consecutive timestep.
They are initialized \SI{1}{m} from each other in a random order and at a random angle between $[-\frac{\pi}{8},\frac{\pi}{8}]$, with zero aligned to West/East.
They take 2D velocity actions which result in the control forces shown as blue lines.
Each agent observes its velocity and obtains the relative position and velocity to the other agent through communication. 

This scenario is designed to undergo a disruption, which manifests itself as an external southbound wind (a force acting in the direction opposite to the agents' desired velocity).
This results in the agents having to exert more work to track a desired velocity.
However, the agents are physically different (see \autoref{fig:wind_scenario}): the triangular agent has an aerodynamically shielding property, which means that, if it flocks in front of the circular agent, it is able to deflect wind and thus reduce the team energy expenditure.
We model the effect of this shielding on the circular agent as proportional to its angular displacement `behind' the triangular agent (in the direction of wind),
with its perceived wind force dipping to \SI{0}{\percent} in case of full alignment.
Conversely, an alignment that keeps them horizontal, or the triangular agent behind, causes both of them to be effected by the same maximum wind.
The agents receive a reward inversely proportional to the total perceived wind, rewarding the team to minimize the total energy needed for the task.
This additional term is simply 0 in the base (windless) version of the task.

In \autoref{fig:wind_reward} and \autoref{fig:wind_snd} we report the reward and SND, respectively,
both for heterogeneous and homogeneous models.
From iteration 0 to 100 no wind is present in the environment.
Both models learn the optimal solution, which is to flock northwards in the formation they are spawned in.
This is because, at this stage, the team has no reason to prefer one formation over the other, and
changing formation would require
sacrifices in velocity-tracking performance.
Furthermore, without wind, the robots cannot benefit from diversity, and thus we observe the heterogeneous model behaving almost homogeneously (SND $\leq 0.5$).

When wind is added to the environment (iteration 100),
the team can now collect additional reward by reducing the 
perceived wind.
The homogeneous model, unable to observe the physical difference of the agents, fails to perform wind shielding and has to employ the same policy for both agents.
On the other hand, the heterogeneous agents, being able to learn diverse policies, learn that they can decrease the wind impacting the team by diverging in behavior and performing wind shielding.
We can see that, during the time window when wind is introduced (100-400), the heterogeneous agents gradually adapt and increase their reward as well as diversity.
We then remove the wind (between iterations 400-600), reverting the POMG to its initial form.
The performance of both paradigms is now the same as before the wind was introduced, with one fundamental difference.
The difference, imperceptible in the reward, is visible in the SND.
The heterogeneous agents have learned to keep the wind shielding formation, and, since this does not have a significant impact on the reward, they retain this skill even during times where it is not needed.
We refer to this as \textit{latent resilience}.

The main advantage of this latent resilience is observed when the wind is reintroduced (iteration 600).
The heterogeneous agents are able to \textit{immediately} obtain the maximum reward thanks to the latent shielding skill learned from the previous appearance of this disturbance.
Furthermore, they are able to do so with less diversity than before.
This is because the agents have learned to act homogeneously in the parts of the task that benefit from it, and thus find the optimal trade-off between homogeneity and heterogeneity through time (as already observed in \autoref{fig:static_size_snd}).

\subsection{Controlling SND: Differential Steering}
\label{sec:control}

\begin{figure}[t]
    \centering
    \includegraphics[width=\linewidth]{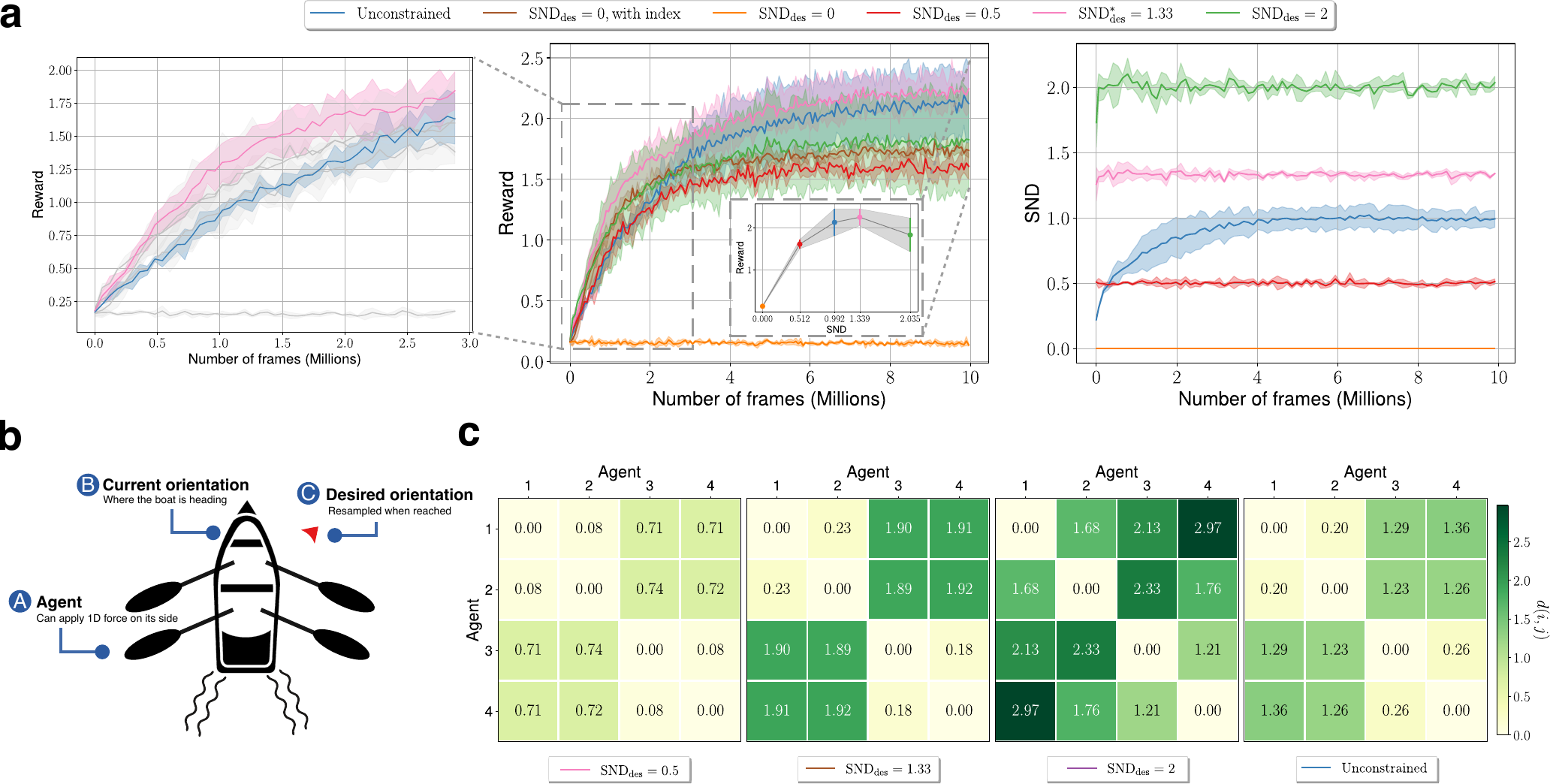}
    \caption{Controlling SND in the \textit{Differential Steering} case study. \textbf{a,} Reward and SND for different levels of controlled diversity and for unconstrained heterogeneous and homogeneous agents. \textbf{b,} The setup of the task. \textbf{c,} behavioral distance matrices for the four heterogeneous runs. 
    The figure shows how we can bootstrap exploration by controlling SND, and how too much or too little SND can be detrimental to the task.
    We report mean and standard deviation over 3 random seeds for each experiment.} 
    \label{fig:control}
\end{figure}

Now that we have seen how SND can be used as an insightful diversity measurement tool, we are interested in showing how it can be used to \textit{control} diversity.
Towards this end, we employ DiCo~\cite{bettini2024controlling}, a diversity control paradigm that uses SND as the reference metric, constraining the multi-agent system to a desired diversity value $\mathrm{SND}_\mathrm{des}$.
We run experiments in BenchMARL~\cite{bettini2024benchmarl} using the MAPPO algorithm.

To illustrate the effects of controlling SND, we create the \textit{Differential Steering} case study, shown in \autoref{fig:control}b.
In this task, we position $\frac{n}{2}$ agents on each side of a rowboat.
Agents can apply a 1D force that affects the orientation of the vessel through a torque.
The effects of this action are flipped for agents on opposite sides, meaning a forward force applied by agents on the right will cause counter-clockwise rotation while the same force applied by agents on the left will cause a clockwise rotation.
The net torque on the boat is given by the sum of the forces applied by all agents. The goal of the agents is to steer the boat to a desired orientation which is randomly resampled when reached.
The agents need to do so by observing only the orientation of the boat, with knowledge of the desired orientation.

The optimal solution requires agents from different sides to take maximally different actions at all times, while agents on the same side take maximally similar actions.
Since the agents do not observe which side they are on, na\"ive homogeneous policies are not able to solve the task.
Given that actions are in the range $[-1,1]$, two agents that converge to Delta distributions on the opposite extremes of this range will have a $W_2$ distance of $2$.
Therefore, we know that the optimal policies for this tasks consist of two behavioral clusters of $\frac{n}{2}$ homogeneous agents, with the clusters being at behavioral distance $2$. Recalling \autoref{prop:redundancy}, we can compute the optimal system SND for this task, replacing $x=2$ (distance between clusters), $n_c=2$ (number of clusters), $n=4$ (the number of agents we use in the experiment), obtaining:
$$
\mathrm{SND}(\mathbf{D}) = x\frac{n(n_c-1)}{n_c(n-1)}= \frac{n}{n-1} = 1.\overline{3}.
$$
We use this derived optimal value $\mathrm{SND}^{*}_\mathrm{des}=1.33$ to control SND in the steering task described above.
We also run experiments with: unconstrained heterogeneous agents (HetMAPPO), homogeneous agents ($\mathrm{SND}_\mathrm{des}=0$), homogeneous agents that observe a one-hot index (to allow them to condition on this to emulate diverse behavior), insufficient controlled diversity ($\mathrm{SND}_\mathrm{des}=0.5$), and excess controlled diversity ($\mathrm{SND}_\mathrm{des}=2$).

From the results (\autoref{fig:control}), we confirm that agents constrained at
$\mathrm{SND}_\mathrm{des} = \mathrm{SND}^{*}_\mathrm{des} = 1.33$
obtain the best reward and are able to bootstrap the exploration phase thanks to the diversity constraint.
Unconstrained heterogeneous agents also arrive at a similar behavior with a comparable SND, but, being not controlled, take longer to converge to this solution.
We also observe that controlling SND to values that are too low ($0.5$) or too high ($2$) can be detrimental to performance as the agents are forced to allocate too little or too much diversity with respect to what is needed for the task.
We also note how, in this case, homogeneous agents that are able to learn multimodal behavior by conditioning on an explicit index, converge to suboptimal performance, probably due to the large discontinuity required to switch between the two behavioral roles.
From the behavioral matrices in \autoref{fig:control}, we observe how $\mathrm{SND}_\mathrm{des}=1.33$ more closely aligns with our theoretical analysis, with an inter-cluster behavioral distance $\approx2$ and an intra-cluster distance $\approx0$.

This simple case study shows how controlling SND to a known optimal value helps in reducing the search space of unconstrained heterogeneous policies, and consequently improve the overall performance of the learned policies. While in this case we were able to compute the optimal $\mathrm{SND}^{*}_\mathrm{des}$ given the POMG, this may not be always possible.
For more complex tasks, where such priors are not available, it is possible to apply the constraint as an inequality (SND$\geq\mathrm{SND}_\mathrm{des}$), which can still provide the aforementioned benefits without constraining the policies to a specific SND value.
For more examples and extensive experiments on controlling SND, we refer the reader to~\citet{bettini2024controlling}.

\section{Discussion}
Behavioral diversity is a valuable skill in collective problems.
As seen from examples above, multi-agent systems that allow agents to specialize and learn unique (and potentially complementary) skills often demonstrate superior resilience towards disturbances.
This work presented a novel framework for capturing this behavioral heterogeneity with a System Neural Diversity (SND) metric.
Our metric is able to measure behavioral diversity among agents with stochastic policies acting in continuous state spaces.
Prior work in designing such diversity metrics overlooked two key properties of interest -- the ability to compare diversity across varying team sizes, and the ability to represent redundancy -- thus capturing only a partial picture of heterogeneity.
We show, theoretically and empirically, that SND functions as a complementary tool that enables these comparisons.
Our evaluations of the metric in a variety of didactic and realistic multi-robot tasks with different system sizes and different heterogeneity requirements establish the representational power of our metric.

The insights we draw from the various tasks indicate that SND provides a means to observe latent properties developed by a heterogeneous system during training, allowing a direct measurement of latent resilience, where other proxies, such as task performance (reward), fail to.
In particular, we highlight the following insights:
\begin{enumerate}
    \item  When training heterogeneous policies, observing $\mathrm{SND}\approx0$ indicates that agents are behaving homogeneously, and that homogeneous training could instead be used to benefit from higher sample efficiency (\autoref{sec:navigation2}).
    \item In tasks where heterogeneous paradigms obtain higher rewards, an $\mathrm{SND} > 0$ acts as confirmation that team heterogeneity enables performance (\autoref{subsec:different_size_passage}). Conversely, when heterogeneous policies add resilience to noise (\autoref{subsec:asymm_payload_passage}), an observed $\mathrm{SND}>0$ explains this diversity while the rewards alone do not.
    \item  In cases where the task undergoes repeated dynamic disruptions, SND is able to expose how heterogeneity allows agents to learn and maintain latent skills that confer resilience (\autoref{subsec:wind}).
    \item Controlling SND helps bootstrapping the exploration phase leading to faster convergence to optimal policies (\autoref{sec:control}).
\end{enumerate}

Finally, while our evaluations suggest that heterogeneous training is a powerful paradigm for multi-agent problems, we caution that this may not always be the case.
In fact, it is possible to construct pathological scenarios where, just like in nature~\cite{mckinney1997extinction}, excessive role specialization within a multi-agent system becomes the cause of failure during unforeseen changes.
For instance, highly specialized agents are not able to dynamically swap roles, and thus, without behavioral redundancy, the system performance may suffer from agent faults.
SND can still play an important role in such cases by aiding the analysis and regulation of a trade-off between specialization and redundancy.

\acks{
This work was supported by Army Research Laboratory (ARL) Distributed and Collaborative Intelligent Systems and Technology (DCIST) Collaborative Research Alliance (CRA) W911NF-17-2-0181 and the European Research Council (ERC) Project 949940 (gAIa). 
We thank Alex McAvoy for reviewing a draft of this manuscript and providing helpful feedback.
}

\newpage
\appendix
\section{Further Related Works}

\label{app:related}

In this section, we present further related works about diversity indices from the biological domain.

Diversity indices are quantitative measures of the number of different species in a community. They are commonly used in ecology and biology  to represent different aspects of diversity such as richness, evenness, and divergence~\cite{tucker2017guide}. They measure the distribution of $n$ elements into $c$ classes, where the proportion of elements in each class $h \in \{1,c\}$ is noted as $p_h$. Most of these indices are special cases of the \textit{true diversity index} or \textit{Hill numbers}~\cite{jost2006entropy}. In particular, Shannon entropy~\cite{shannon1948mathematical} is equivalent to the logarithm of the true diversity of order 1 and takes the following form:

\begin{equation}
    E = -\sum_{h=1}^c p_h \log_2 p_h.
    \label{eq:shannon}
\end{equation}
This index is adopted across a wide range of domains as it weights each class by its proportional abundance.

Despite the wide adoption of diversity indices, they are not directly applicable to measuring behavioral diversity in MARL.
This is because they rely on a predefined number of classes or species, while the multi-agent systems we consider reside in a continuous time-varying behavioral (i.e., policy) space. 
\section{Sample efficiency of homogeneous training}

\label{app:hom_train}
\begin{figure}[t]
    \centering
    \begin{subfigure}{0.5\linewidth}
       \includegraphics[width=\columnwidth]{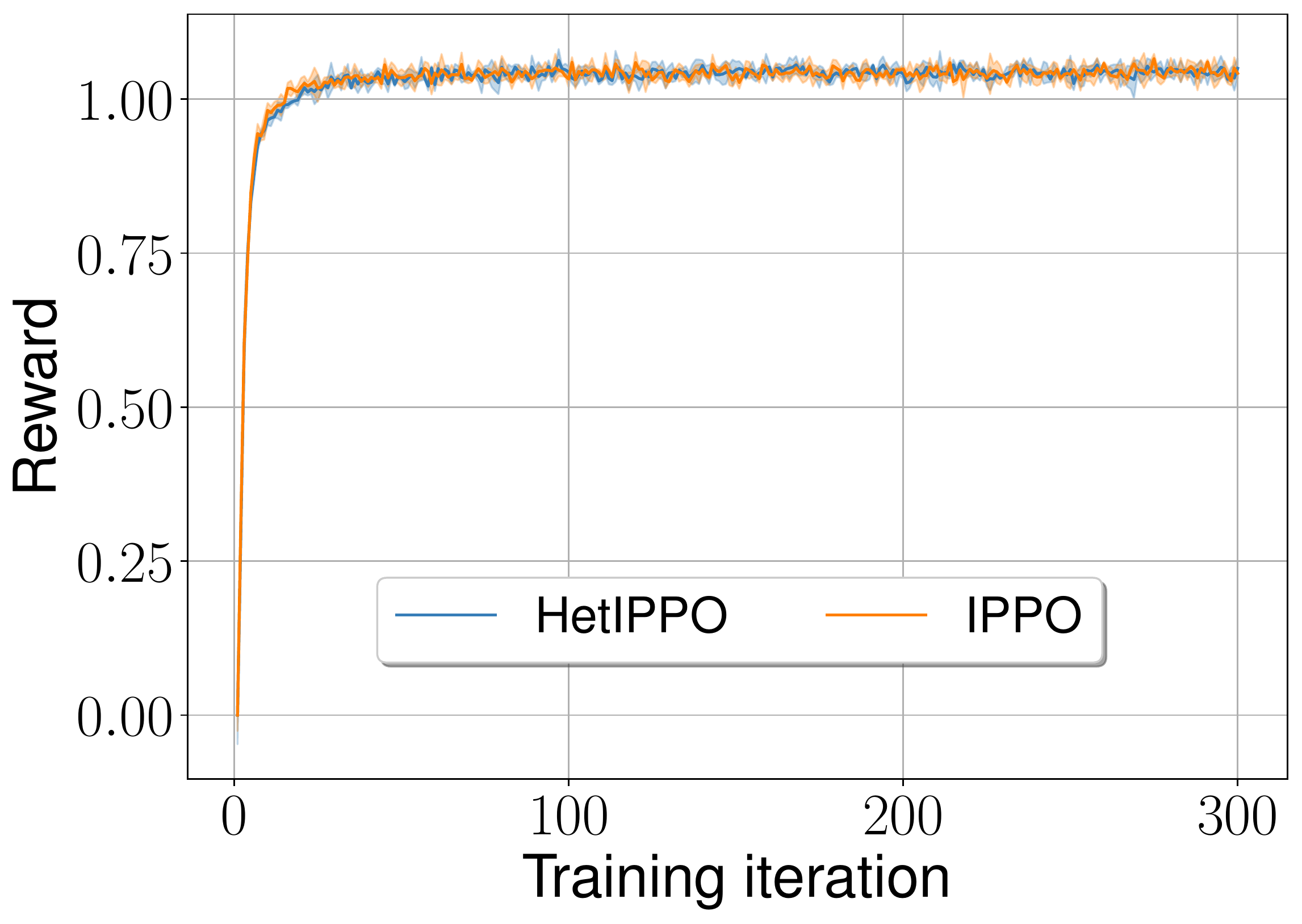}
       \caption{Reward}
        
    \end{subfigure}\hfill
    \begin{subfigure}{0.5\linewidth}
        \includegraphics[width=\columnwidth]{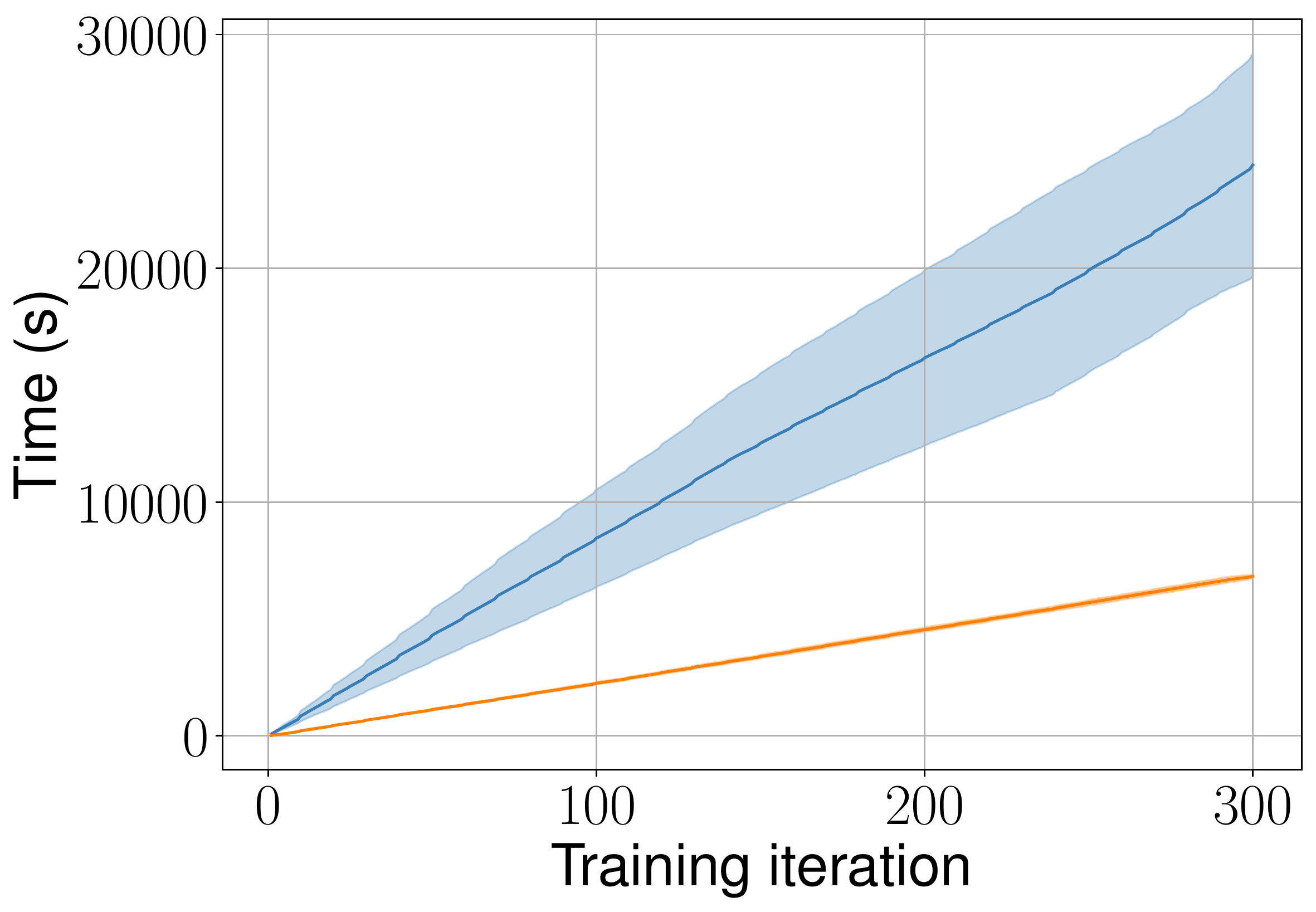}
        \caption{Cumulative training time}
    
    \end{subfigure}\hfill
    \caption{Comparison of homogeneous (IPPO) and heterogeneous (HetIPPO) training in the \textit{Multi-Agent Goal Navigation} scenario with 1 goal, where heterogeneous models have been shown to have approximately 0 SND. In this case, we can observe that homogeneous training should be preferred as a more time-efficient solution thanks to its higher sample efficiency.  We report mean and standard deviation over 3 random seeds for each experiment. The values are measured over 300 training iterations each performed over 600 episodes of experience.}
    \label{fig:multi_goal_hom_v_het}
\end{figure}

In this section, we are interested in demonstrating the benefits of homogeneous training in the \textit{Multi-Agent Goal Navigation}, when all agents are assigned the same goal and SND approaches 0.
In such a case, the metric acts an important diagnostic tool, suggesting that a homogeneous training strategy should be preferred in subsequent experiment iterations in order to benefit from parameter sharing and increased sample efficiency.

To demonstrate this, we run an additional evaluation. We compare homogeneous and heterogeneous training. In \autoref{fig:multi_goal_hom_v_het} we show that both paradigms obtain the same performance, with the homogeneous model being significantly more time-efficient.
\section{Experimental setup}

Simulations are performed in the VMAS~\cite{bettini2022vmas} simulator.
Agents are trained using the (Het)GPPO model~\cite{bettini2023hetgppo} with a fully-connected graph topology.
We refer to the non-communication version of (Het)GPPO as Het(IPPO).
Agent policies output a 2D continuous action distribution for each observation.
The distribution is parameterized using two univariate Gaussians (one for each dimension).
Training is performed in RLlib~\cite{liang2018rllib} using PyTorch~\cite{paszke2019pytorch} and a multi-agent implementation of the PPO algorithm~\cite{blumenkamp2021emergence}. The training parameters used are shown in \autoref{tab:rl_params}.

The code for the experiments used in this paper is publicly available at \url{https://github.com/proroklab/HetGPPO}.

\begin{table}[t]
    \centering
    \caption{Training parameters for all evaluations.}
    \label{tab:rl_params}
    \begin{tabular}{lc|cc}
    \multicolumn{2}{c}{Training} & \multicolumn{2}{c}{PPO} \\ \toprule
    Batch size & 60000  & $\epsilon$ & 0.2 \\ 
    Minibatch size & 4096 & $\gamma$ & 0.99  \\ 
    SDG Iterations & 40 & $\lambda$ & 0.9 \\ 
    \# Workers &  5 & Entropy coeff & 0 \\
    \# Envs per worker &  50 & KL coeff & 0.01 \\
    Learning rate & 5e-5 & KL target & 0.01\\
    \end{tabular}
\end{table}

\newpage
\bibliography{references}

\end{document}